%% file: main.tex
\documentclass[a4paper,11pt]{article}
\usepackage{jheppub} 

\usepackage{physics}
\usepackage{amsmath}
\usepackage{bm}
\usepackage{bbm}
\usepackage{multicol}
\usepackage{slashed}
\usepackage{multirow}
\usepackage{multicol}
\usepackage{cleveref}
\usepackage{todonotes}
\usepackage[most]{tcolorbox}
\usepackage{float}
\usepackage{mathtools}
\usepackage[normalem]{ulem}

\usepackage{datatool, filecontents}
\DTLsetseparator{,}

\DTLloaddb[noheader, keys={thekey,thevalue,theerror,theexp}]{fitdata}{Fits/fitpars.formatted.csv}
\newcommand{\fitval}[1]{\DTLfetch{fitdata}{thekey}{#1}{thevalue}}
\newcommand{\fiterr}[1]{\DTLfetch{fitdata}{thekey}{#1}{theerror}}
\newcommand{\fitexp}[1]{\DTLfetch{fitdata}{thekey}{#1}{theexp}}

\newcommand{\fitvalerrNoexp}[1]{\fitval{#1}(\fiterr{#1})}

\input{Definitions}

\title{Exploratory calculation of the rare hyperon decay \boldmath{$\Sigma^+ \to p \ell^+ \ell^-$} from lattice QCD}

\author[a]{Felix Erben}
\author[b]{Vera G\"ulpers}
\author[b]{Maxwell T. Hansen}
\author[c,b]{Raoul Hodgson}
\author[b]{Antonin Portelli}

\affiliation[a]{Department of Theoretical Physics, CERN, 1211 Geneva 23, Switzerland}
\affiliation[b]{Higgs Centre for Theoretical Physics, School of Physics and Astronomy, The University of Edinburgh, Edinburgh EH9 3FD, United Kingdom}
\affiliation[c]{Deutsches Elektronen-Synchrotron DESY, Platanenallee 6, 15738 Zeuthen, Germany}

\emailAdd{felix.erben@cern.ch}
\emailAdd{vera.guelpers@ed.ac.uk}
\emailAdd{maxwell.hansen@ed.ac.uk}
\emailAdd{raoul.hodgson@desy.de}
\emailAdd{antonin.portelli@ed.ac.uk}

\abstract{\input{Sections/Abstract}}

\begin{document}
\begin{filecontents*}{Fits/fitpars.formatted.csv}
m_Nuc,0.65286880,0.00459460,0
m_Sig,0.76457250,0.00364464,0
q2,-0.06479763,0.00050730,0
q2Max,0.01247771,0.00097263,0
RH_Eye_a,-25.4,37.1,0
RH_Eye_c,-3.5,6.9,-2
RH_Eye_muT,-0.64,0.95,-1
RH_Eye_muZ,-1.79,3.07,-2
RH_Eye_rho_muT,0.13,0.68,-1
RH_Eye_rho_muZ,0.71,1.98,-2
RH_Eye_sig_muT,-0.77,0.74,-1
RH_Eye_sig_muZ,-2.49,2.40,-2
RH_NE_a,-6.2,8.3,0
RH_NE_c,-1.6,1.7,-2
RH_NE_dt12_a,-5.7,2.3,0
RH_NE_dt12_c,-1.7,0.5,-2
RH_NE_dt12_muT,-0.27,0.07,-1
RH_NE_dt12_muZ,-0.34,0.19,-2
RH_NE_dt12_rho_muT,2.59,0.18,-1
RH_NE_dt12_rho_muZ,-3.98,0.28,-2
RH_NE_dt12_sig_muT,-2.86,0.16,-1
RH_NE_dt12_sig_muZ,3.64,0.21,-2
RH_NE_meas_a,-2.7,5.4,0
RH_NE_meas_c,1.1,1.1,-2
RH_NE_meas_muT,0.14,0.17,-1
RH_NE_meas_muZ,-0.28,0.42,-2
RH_NE_meas_rho_muT,2.61,0.30,-1
RH_NE_meas_rho_muZ,-3.92,0.54,-2
RH_NE_meas_sig_muT,-2.47,0.23,-1
RH_NE_meas_sig_muZ,3.64,0.35,-2
RH_NE_muT,-0.26,0.23,-1
RH_NE_muZ,-0.39,0.70,-2
RH_NE_rho_muT,2.81,0.33,-1
RH_NE_rho_muZ,-3.63,0.72,-2
RH_NE_sig_muT,-3.07,0.26,-1
RH_NE_sig_muZ,3.24,0.55,-2
RH_NE_sub_a,10.2,35.6,0
RH_NE_sub_c,10.9,7.2,-2
RH_NE_sub_muT,1.60,1.01,-1
RH_NE_sub_muZ,0.11,2.93,-2
RH_NE_sub_rho_muT,0.48,0.95,-1
RH_NE_sub_rho_muZ,6.48,3.75,-2
RH_NE_sub_sig_muT,1.12,1.21,-1
RH_NE_sub_sig_muZ,-6.38,3.78,-2
RH_Tot_a,-27.0,39.0,0
RH_Tot_c,-2.7,9.4,-2
RH_Tot_muT,-0.53,1.33,-1
RH_Tot_muZ,-1.98,3.21,-2
RH_Tot_rho_muT,3.41,1.14,-1
RH_Tot_rho_muZ,-2.80,2.50,-2
RH_Tot_sig_muT,-3.94,0.93,-1
RH_Tot_sig_muZ,0.82,1.96,-2
\end{filecontents*}

\begin{flushright}
CERN-TH-2025-055 \\
DESY-25-057
\end{flushright}
\vspace{-1.5cm}

\maketitle
\flushbottom
\clearpage

\section{Introduction}
\input{Sections/Intro}

\section{Lattice methodology}
\label{sec:lat-meth}
\input{Sections/Lattice_Method}

\section{Lattice setup}
\label{sec:lat-setup}
\input{Sections/Lattice_Setup}

\section{Numerical results}
\label{sec:results}
\input{Sections/Results}

\section{Summary and outlook}
\label{sec:summary}
\input{Sections/Summary}

\acknowledgments
The authors thank the members of the RBC and UKQCD Collaborations for helpful discussions and suggestions. This work used the DiRAC Extreme Scaling service Tursa at the University of Edinburgh, managed by the Edinburgh Parallel Computing Centre on behalf of the STFC DiRAC HPC Facility (\href{www.dirac.ac.uk}{www.dirac.ac.uk}). The DiRAC service at Edinburgh was funded by UKRI and STFC capital funding and STFC operations grants. DiRAC is part of the UKRI Digital Research Infrastructure. F.E.~has received funding from the European Union's Horizon Europe research and innovation programme under the Marie Sk\l{}odowska-Curie grant agreement No.~101106913. V.G., M.T.H and A.P.~are supported in part by UK STFC grants ST/X000494/1 and ST/T000600/1. F.E., V.G., R.H and A.P.~also received funding from the European Research Council (ERC) under the European Union's Horizon 2020 research and innovation programme under grant agreement No.~757646. A.P.~additionally received additional funding under grant agreement No.~813942. M.T.H.~is further supported by UKRI Future Leaders Fellowship MR/T019956/1.

\appendix
\section{Covariance matrices}
\label{ap:cov_mat}
\input{Sections/Ap_Cov_Mat}

\section{Discrete time}
\label{ap:discrete_time}
\input{Sections/Ap_Discrete_Time}

\clearpage
\section{Four-point diagrams}
\label{ap:4pt_diagrams}
\input{Sections/AP_fourptdiagrams}
\clearpage

\section{Negative parity}
\label{ap:negative_parity}
\input{Sections/Ap_Negative_Parity}

\section{Alternative methods}
\label{ap:alt_methods}
\input{Sections/Ap_Alt_Methods}

\bibliographystyle{JHEP}
\bibliography{Refs}

\end{document}

%% file: Definitions.tex
\def\inf {\ensuremath{\infty}}

\newcommand{\vect}[1]{\ensuremath{\bm{#1}}}

\newcommand{\fm}{\text{fm}}
\newcommand{\Mev}{\text{MeV}}
\newcommand{\Gev}{\text{GeV}}

\newcommand{\cov}{\text{cov}}

%% file: Sections/Abstract.tex
The rare hyperon decay $\Sigma^+ \to p \ell^+ \ell^-$ is a flavour-changing neutral current process mediated by an $s \to d$ transition that occurs only at loop level within the Standard Model. Consequently, this decay is highly suppressed, making it a promising avenue for probing potential new physics. While phenomenological calculations have made important progress in predicting the decay amplitude, there remains a four-fold ambiguity in the relevant transition form factors that prevents a unique prediction for the branching fraction and angular observables. Fully resolving this ambiguity requires a first-principles Standard-Model calculation, and the recent observation of this process using LHCb Run 2 data reinforces the timeliness of such a calculation. In this work, we present the first lattice-QCD calculation of this decay, performed using a 2+1-flavour domain-wall fermion ensemble with a pion mass of $340 \, \Mev$. At a small baryon source-sink separation, we observe the emergence of a signal in the relevant baryonic four-point functions. This allows us to determine the positive-parity form factors for the rare hyperon decays from first-principles, albeit with large statistical and systematic uncertainties.

%% file: Sections/Intro.tex
The rare hyperon decay $\Sigma^+ \to p \ell^+ \ell^-$ is a flavour-changing neutral current (FCNC) process that is heavily suppressed within the Standard Model (SM) and is therefore sensitive to new physics.
Evidence for this decay with muonic final states has been obtained by both the HyperCP and LHCb experiments in refs.~\cite{HyperCPCollaboration2005Evidencemu} and \cite{LHCb:2017rdd} respectively, which give a combined branching fraction measurement of \cite{ParticleDataGroup:2024cfk}
\begin{align}
    \mathcal{B}(\Sigma^+ \to p \mu^+ \mu^-)_\text{PDG}=\left(2.4^{+1.7}_{-1.3}\right)\times 10^{-8}\,.
\end{align}
However, the LHCb collaboration have recently presented the first observation of this decay above $5\sigma$ statistical significance in ref.~\cite{LHCb-CONF-2024-002} resulting from both an increased dataset and improvements to the trigger system \cite{Dettori:2297352}, and a full analysis of the branching fraction with this updated data set is presented in ref.~\cite{LHCb:2025evf}, 
\begin{align}
    \mathcal{B}(\Sigma^+ \to p \mu^+ \mu^-)_\text{LHCb 2025}= \left(1.08 {\pm 0.17} \right)\times 10^{-8}\,.
\end{align}
With this updated dataset, determinations of other quantities such as the differential branching ratio, forward-backward asymmetry and CP violation may also be possible and future publications on these results are envisaged.

References~\cite{He2005DecayModel,He2018DecayPmu+mu-,Roy:2024hqg} have provided the existing SM prediction of the four hadronic form factors governing this decay (denoted by $a$,$b$,$c$ and $d$) based on a combination of techniques including baryon ChPT and vector meson dominance models, which make use of experimental inputs. These studies found that the decay is dominated by the long-distance intermediate virtual photon process $\Sigma^+ \to p \gamma^\ast \to p \ell^+ \ell^-$. The experimental data used for the published determinations include the $\Sigma^+ \to p \gamma$ and $\Sigma^+ \to N \pi$ decay amplitudes, which were both recently updated by the BESIII experiment in refs.~\cite{BESIII:2023fhs} and \cite{BESIII:2023sgt} respectively. Since the experimental data do not fully constrain the relevant form factors, there remains a four-fold ambiguity in the SM prediction, corresponding to various branching fractions in the range
\begin{align}
    1.2 \times 10^{-8} < \mathcal{B}(\Sigma^+ \to p \mu^+ \mu^-)_\text{SM} < 7.8 \times 10^{-8} \,.
    \label{eq:SM-pred}
\end{align}
However, the recent experimental measurement \cite{LHCb:2025evf} strongly favours the smallest of these predicted values $\mathcal{B} = (1.2 \pm 0.1) \times 10^{-8}$. It has also been shown in ref.~\cite{Roy:2024hqg} that with the additional measurement of the $\Sigma^+ \to p \gamma$ spin-projections and photon polarisation, the form factors can be further constrained down to just a two-fold ambiguity. However, such an experimental measurement has yet to be performed.

Due to these remaining ambiguities in the form factors, it would be highly beneficial to have a first-principles computation using lattice QCD. If sufficiently precise, the results of such a calculation could be used in combination with the existing phenomenological calculations to help fully determine all four form factors. For example, a lattice determination of the sign of $\Re a$ would reduce from a four-fold to a two-fold ambiguity, with the latter orthogonal to the reduction that could be obtained from the $\Sigma^+ \to p \gamma$ polarisation measurements.
This would then reduce the range of \eqref{eq:SM-pred} to
a single value for the branching fraction. In addition, because lattice QCD is systematically improvable, such a calculation may eventually be able to provide a fully ab initio calculation of this decay process without the need for any experimental or phenomenological input (beyond the hadron masses used to set the scale and quark mass inputs).

In addition, the $a$ and $b$ form factors are relevant for the SM prediction of the real photon emission process $\Sigma^+ \to p \gamma$ for which there is current interest due to unresolved tensions between theoretical predictions and experimental observations in the set of processes referred to as weak radiative hyperon decays (WRHDs) \cite{Shi:2025xkp}. There are six WRHD channels: $\Sigma^+ \to p \gamma$, $\Lambda/\Sigma^0 \to n \gamma$, $\Xi^0 \to \Lambda/\Sigma^0 \gamma$ and $\Xi^- \to \Sigma^- \gamma$, all of which can be investigated via lattice QCD using the framework presented by this set of authors in ref.~\cite{Erben:2022tdu} and used in this work.

The comprehensive framework for a full lattice QCD computation of the rare hyperon decay described in ref.~\cite{Erben:2022tdu}, which includes methods for relating finite-volume matrix elements to the physical observable, builds upon earlier foundational work~\cite{Christ2015ProspectsDecays, Briceno:2019opb}. In this paper, we present the first exploratory calculation applying this formalism at an unphysically heavy pion mass, focusing on the positive-parity hadronic form factors associated with this decay process. We begin with a brief review of the formalism in \cref{sec:lat-meth}, followed by a detailed description of the specific setup used in our calculation in \cref{sec:lat-setup}. The numerical results and their discussion are presented in \cref{sec:results}, and we conclude and given an outlook in \cref{sec:summary}.

%% file: Sections/Lattice_Method.tex
In this section we give a brief overview of the procedure of extracting the rare hyperon decay from lattice QCD. For a comprehensive description see ref.~\cite{Erben:2022tdu}.

\subsection{Amplitude in Minkowski space-time}
The goal of this project is to compute the long-distance decay amplitude for the dominant process $\Sigma^+ \to p \gamma^\ast$, which is given by
\begin{align}
\label{eq:Amp_def}
    \mathcal{A}^{rs}_\mu & = \int d^4x \bra{p(\vect{p}),r} T \left\{ H_W(x) J_\mu(0) \right\} \ket{\Sigma^+(\vect{k}),s} \,,
\end{align}
where $r,s$ are spin projection labels, $\vect{k}$ and $\vect{p}$ are the three-momenta of the initial and final states respectively, $J_\mu$ is the electromagnetic current, and $H_W(x)$ is the effective weak $s \to d$ Hamiltonian density constructed from four-quark operators \cite{Buchalla1995WeakLogarithms}
\begin{align}
    H_W & = \frac{G_F}{\sqrt{2}} V_{us} V_{ud}^\ast \left[ C_1 (Q_1^u-Q_1^c) + C_2 (Q_2^u-Q_2^c) + \dots \right]  \,, \label{eq:weakH}\\
    Q_1^q & = [\bar{s} \gamma_\mu (1-\gamma_5) d] \, [\bar{q} \gamma^\mu (1-\gamma_5) q] \label{eq:Q1} \,,\\[5pt]
    Q_2^q & = [\bar{s} \gamma_\mu (1-\gamma_5) q] \, [\bar{q} \gamma^\mu (1-\gamma_5) d] \label{eq:Q2}\, ,
\end{align}
where the quark flavour $q$ can be either an up or charm quark as shown. $C_i$ are the Wilson coefficients of the effective interaction vertices, and the ellipsis indicates additional operators that are suppressed relative to these first two \cite{Erben:2022tdu}, and are therefore not considered in this work.

This amplitude can be written with a form factor decomposition
\begin{align}
    \mathcal{A}^{rs}_\mu & = -i \, \overline{u}_p^r(\vect{p}) \left[ i \sigma_{\nu \mu} q^{\nu} (a(q^2)+b(q^2) \gamma_5) + (q^2 \gamma_\mu - q_\mu \slashed{q}) (c(q^2)+d(q^2) \gamma_5) \right] u_\Sigma^s(\vect{k}) \,,
\end{align}
where $q=k-p$ is the four-momentum transfer, and the form factors $a(q^2),b(q^2),c(q^2),d(q^2)$ are scalar functions and therfore depend only on the squared momentum transfer $q^2$. The outermost factors, $u_\Sigma$ and $\bar{u}_p$, are the initial and final state spinors, respectively.

\begin{figure}
    \centering
    \includegraphics[width=0.8\linewidth]{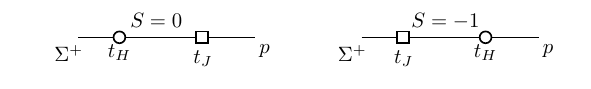}
    \caption{The two time orderings of the weak Hamiltonian at time $t_H$ and the electromagnetic current at time $t_J$. The strangeness quantum number of the intermediate states are also indicated.}
    \label{fig:timeorderings}
\end{figure}

Finally, this amplitude can be written in terms of two spectral functions $\rho$ and $\sigma$ that correspond to the two time orderings of the intermediate operators $H_W$ and $J_\mu$, depicted schematically in \cref{fig:timeorderings},
\begin{align}
\label{eq:Amp_spectral}
    \mathcal{A}^{rs}_\mu & = \lim_{\epsilon\to 0^+} -i \int_0^\inf d \omega \left( \frac{\rho^{rs}_\mu(\omega)}{\omega-E_\Sigma(\vect{k})- i \epsilon} + \frac{\sigma^{rs}_\mu(\omega)}{\omega-E_p(\vect{p})- i \epsilon} \right) \,.
\end{align}
This form is especially important for the extraction of the amplitude on the lattice. The spectral functions are defined by
\begin{align}
    \rho^{rs}_\mu(\omega) & = \int\mathllap{\sum_\alpha} \frac{\delta(\omega-E_\alpha(\vect{k}))}{2 E_\alpha(\vect{k})} \bra{p(\vect{p}),r} J_\mu(0) \ket{E_\alpha(\vect{k})} \bra{E_\alpha(\vect{k})} H_W(0) \ket{\Sigma^+(\vect{k}),s} \,, \\
    \sigma^{rs}_\mu(\omega) & = \int\mathllap{\sum_\beta} \frac{\delta(\omega-E_\beta(\vect{p}))}{2 E_\beta(\vect{p})} \bra{p(\vect{p}),r} H_W(0) \ket{E_\beta(\vect{p})} \bra{E_\beta(\vect{p})} J_\mu(0) \ket{\Sigma^+(\vect{k}),s} \,,
\end{align}
where the sum/integral over intermediate states runs over all energy eigenstates with baryon number 1, and strangeness $S=0$ or $-1$ for $\rho$ and $\sigma$ respectively.

\subsection{Amplitude from Euclidean correlators}
The Euclidean finite-volume equivalent of \cref{eq:Amp_def} is the four-point function
\begin{align}
\label{eq:4pt_def}
    \Gamma^{(4)}_\mu(t_p,t_H,t_J) & = \int d^3 \, \vect{x} \langle \psi_p(t_p,\vect{p}) H_W(t_H,\vect{x}) J_\mu(0) \overline\psi_\Sigma(t_\Sigma,\vect{k}) \rangle \,,
\end{align}
where $\psi_p$ and $\overline{\psi}_\Sigma$ are unpolarised interpolators (in the time-momentum representation) for the proton and $\Sigma^+$ respectively. For simplicity we exploit time translational invariance and fix the electromagnetic current time to $t_J=0$.
It should be noted that in practice it is generally advantageous to also project the electromagnetic current to definite three-momentum, in which case the momentum is over-constrained and therefore an additional factor of the volume is present that should be removed.

As is described in \cite{Erben:2022tdu}, instead of using objects that live in spin-polarisation space, $X^{rs}$ (where $X \in \{\mathcal{A}_\mu , \rho_\mu, \sigma_\mu , \dots \}$), we can work with objects that are Dirac matrix valued, $\widetilde{X}$, by factoring out the external spinors
\begin{align}
    X^{rs} = \overline{u}_p^r(\vect{p}) \, \widetilde{X} \, u_\Sigma^s(\vect{k}) \,.
\end{align}
With this definition, $\widetilde{X}$ is not a unique quantity, however by defining the Euclidean projectors
\begin{align}
    \mathbb{P}_\Sigma(\vect{k}) = \frac{-i \slashed{k} + m_\Sigma}{2 m_\Sigma} \hspace{2em} \text{and} \hspace{2em}
    \mathbb{P}_p(\vect{p}) = \frac{-i \slashed{p} + m_p}{2 m_p} \,,
\end{align}
we can construct a unique object
\begin{align}
    \mathbb{P}_p(\vect{p}) \widetilde{X} \mathbb{P}_\Sigma(\vect{k}) = \frac{1}{4 m_p m_\Sigma} \sum_{rs} u_p^r(\vect{p}) X^{rs} \overline{u}_\Sigma^s(\vect{k}) \,,
\end{align}
which is the form we use for all relevant objects in this paper. For notational simplicity we drop the three-momentum dependence of these projectors (e.g. $\mathbb{P}_p\widetilde{X} \mathbb{P}_\Sigma$) unless they are different from $\vect{k}$ and $\vect{p}$ for $\mathbb{P}_\Sigma$ and $\mathbb{P}_p$ respectively.

By examining the spectral decomposition of the four-point function \cref{eq:4pt_def} and assuming ground state dominance of the external states ($t_\Sigma \ll 0,t_H \ll t_p$), it can be written
\begin{align}
    \Gamma^{(4)}_\mu(t_p,t_H,t_\Sigma) & = Z_{\Sigma p}(t_p, t_\Sigma) \, \int_0^\inf d \omega \left\{ \begin{array}{ccc}
        \mathbb{P}_p \widetilde{\rho}_\mu(\omega)_L \mathbb{P}_\Sigma \, e^{-(E_\Sigma(\vect{k})-\omega)t_H} & \text{for} & t_H < 0 \\
        \mathbb{P}_p \widetilde{\sigma}_\mu(\omega)_L \mathbb{P}_\Sigma \, e^{-(\omega - E_p(\vect{p}))t_H} & \text{for} & t_H > 0
    \end{array} \right. \,,
\end{align}
where the factor $Z_{\Sigma p}(t_p, t_\Sigma)$ contains information about the creation, propagation and annihilation of the external $\Sigma^+$ and $p$ states
\begin{align}
    Z_{\Sigma p}(t_p, t_\Sigma) = Z_\Sigma Z_p^\ast \frac{m_\Sigma}{E_\Sigma(\vect{k})} \frac{m_p}{E_p(\vect{p})} e^{E_\Sigma(\vect{k})t_\Sigma} e^{-E_p(\vect{p}) t_p} \,,
\end{align}
and can be constructed from energies and overlap factors from the ground state of two-point functions
\begin{align}
\label{eq:2pt_def}
    \Gamma^{(2)}_B(t) & = \langle \psi_B(t_p,\vect{l}) \overline\psi_B(0,\vect{l}) \rangle = L^3 \sum_n |Z_n(\vect{l})|^2 \frac{m_n}{E_n(\vect{l})} \mathbb{P}_n(\vect{l}) e^{-E_n(\vect{l})t} \,,
\end{align}
where $B \in \{p,\Sigma^+\}$ and the three-momentum $\vect{l}=\vect{k}$ or $\vect{p}$. The volume factor $L^3$ comes from the over constraint of the momentum since both operators are projected to definite momentum. Removing the $Z_{\Sigma p}$ factor gives the amputated four-point function
\begin{align}
    \hat\Gamma^{(4)}_\mu(t_H) & = \frac{\Gamma^{(4)}_\mu(t_p,t_H,t_\Sigma)}{Z_{\Sigma p}(t_p, t_\Sigma)} \,.
\end{align}

The spectral functions $\widetilde{\rho}_L$ and $\widetilde{\sigma}_L$ are different to those in \cref{eq:Amp_spectral} in that they are finite-volume objects, which has the effect that instead of being a continuous function of $\omega$ as in the infinite volume, they are described by a sum of Dirac delta functions at finite-volume energies $E_n$
\begin{align}
    \rho^{rs}_\mu(\omega)_L & = \sum_{n} \frac{\delta(\omega - E_n)}{2 E_n} \bra{p(\vect{p}),r} J_\mu(0) \ket{E_n}_L \bra{E_n} H_W(0) \ket{\Sigma(\vect{k}),s}_L \,, \\
    \sigma^{rs}_\mu(\omega)_L & = \sum_m \frac{\delta(\omega - E_m)}{2 E_m} \bra{p(\vect{p}),r} H_W(0) \ket{E_m}_L \bra{E_m} J_\mu(0) \ket{\Sigma(\vect{k}),s}_L \,,
\end{align}
and the subscript $L$ on the matrix elements indicates that they are finite volume objects.

As is described in ref. \cite{Erben:2022tdu}, we construct the integrated correlator as
\begin{align}
\label{eq:Int_all}
    I_\mu(T_a,T_b) = -i \int_{-T_a}^{T_b} dt_H \hat{\Gamma}^{(4)}_\mu(t_H) & = -i \int_0^\inf d\omega \left[ \mathbb{P}_p\widetilde{\rho}_\mu(\omega)_L \mathbb{P}_\Sigma \frac{1-e^{-(\omega-E_\Sigma(\vect{k}))T_a}}{\omega - E_\Sigma(\vect{k})} \right. \\
    & \hspace{6em} \left. + \mathbb{P}_p \widetilde{\sigma}_\mu(\omega)_L \mathbb{P}_\Sigma \frac{1-e^{-(\omega-E_p(\vect{p}))T_b}}{\omega - E_p(\vect{p})} \right] \,. \nonumber
\end{align}
The integrated correlator is related to the finite-volume estimator of the target amplitude \cref{eq:Amp_def} by removing the exponential terms dependent on $T_a$ and $T_b$, following an approach analogous to lattice calculations of the rare kaon decay~\cite{Christ2016FirstKtopiell+ell-,Boyle:2022ccj}. It is advantageous to split the full integral into two independent integrals, one over each time ordering
\begin{align}
    I_\mu(T_a,T_b) & = I^\rho_\mu(T_a) + I^\sigma_\mu(T_b) \,, \\
\label{eq:Int_rho}
    I^\rho_\mu(T_a) & = -i \int_{-T_a}^0 dt_H \hat{\Gamma}^{(4)}_\mu(t_H) = -i \int_0^\inf d\omega \, \mathbb{P}_p \widetilde{\rho}_\mu(\omega)_L \mathbb{P}_\Sigma \frac{1-e^{-(\omega-E_\Sigma(\vect{k}))T_a}}{\omega - E_\Sigma(\vect{k})} \,, \\
\label{eq:Int_sig}
    I^\sigma_\mu(T_b) & = -i \int_0^{T_b} dt_H \hat{\Gamma}^{(4)}_\mu(t_H) = -i \int_0^\inf d\omega \, \mathbb{P}_p \widetilde{\sigma}_\mu(\omega)_L \mathbb{P}_\Sigma \frac{1-e^{-(\omega-E_p(\vect{p}))T_b}}{\omega - E_p(\vect{p})} \,.
\end{align}
The advantages of this separation for practical lattice simulations will be discussed later in \cref{sec:Extraction_Methods,ap:cov_mat}.

The integral over time requires a continuum four-point function rather than with a finite lattice spacing at which measurements are obtained. Therefore this integral must be replaced with a discrete sum, to which there is no unique discretisation. \Cref{ap:discrete_time} discusses some possible choices of discretisation and the cut-off effects they induce. For the purposes of this work, we use the trapezium rule integrator which gives the integrated correlators
\begin{align}
    I^\rho_\mu(T_a>0) = & -i \, a \left( \frac12 \hat{\Gamma}^{(4)}_\mu(0) + \sum_{n=1}^{T_a/a-1} \hat{\Gamma}^{(4)}_\mu(-an) + \frac12 \hat{\Gamma}^{(4)}_\mu(-T_a) \right) \,, \\
    I^\sigma_\mu(T_b>0) = & -i \, a \left( \frac12 \hat{\Gamma}^{(4)}_\mu(0) + \sum_{n=1}^{T_b/a-1} \hat{\Gamma}^{(4)}_\mu(an) + \frac12 \hat{\Gamma}^{(4)}_\mu(T_b) \right)\,,
\end{align}
which introduces $O(a^2)$ cut-off effects from the summation.

As is explained in detail in ref. \cite{Erben:2022tdu}, the integrated correlation functions in general contain growing exponential terms in $T_a$ or $T_b$ that occur when there exist intermediate states with energies lower than the external states. In addition, there are power-like finite-volume corrections that arise when these states below threshold are finite-volume multiparticle states. At physical pion mass the growing intermediate states are the single proton and nucleon-pion ($N\pi$) states, the latter of which contribute the power-like finite volume corrections. However, for the exploratory calculation presented here, the $N\pi$ states are above threshold meaning the amplitude and the finite-volume estimator are simply equivalent up to exponentially suppressed corrections, $\mathcal{A}_\mu = F_\mu + O(e^{-m_\pi L})$, that we neglect in this work. Therefore we do not distinguish between the two and will refer to them as the amplitude for the remainder of this paper.

%% file: Sections/Lattice_Setup.tex
For this exploratory calculation, we use an ensemble with lattice spacing $a=0.11 \, \fm = (1785 \, \Mev)^{-1}$ and $2+1$ flavours of dynamical domain-wall fermions from the RBC-UKQCD collaboration \cite{RBC:2010qam,RBC:2012cbl,RBC:2014ntl}. The lattice extent is $24^3 \times 64 (\times 16)_{L_s}$ where the final number indicates the extend in the fifth dimension required by the domain-wall formalism. We measure all quantities on 70 decorrelated configurations. The relevant hadron masses on this ensemble are summarised in \cref{tab:had_masses}.  Due to these unphysical masses, the lowest $N \pi$ state has larger energy than the $\Sigma^+$ state, and therefore the multiparticle states are above threshold, leading to no power-like finite-volume effects to be accounted for, and only the single proton state contributing a growing intermediate exponential in eq.~\eqref{eq:Int_rho}. In addition to the 3 dynamical light/strange quarks on this ensemble, the weak Hamiltonian also contains a charm quark which we incorporate only in the valence measurements. For this exploratory calculation we use an unphysically light charm mass corresponding to the connected $\eta_c$ meson of mass of $1819(1) \, \Mev$.
\begin{table}[]
    \centering
    \begin{tabular}{c|c|c|c}
        $m_\pi [\Mev]$ & $m_K [\Mev]$ & $m_p [\Mev]$ & $m_\Sigma [\Mev]$ \\
        \hline
        $340(1)$ & $594(1)$ & $1180(12)$ & $1375(9)$
    \end{tabular}
    \caption{Masses of the lowest light and light-strange hadrons relevant for this work. The values of meson masses are taken from \cite{RBC:2010qam}, while the baryon masses are measured as part of this work.}
    \label{tab:had_masses}
\end{table}

The interpolators used in the calculation of the correlation functions are Coulomb gauge-fixed Gaussian smeared ones
\begin{align}
    \psi_p (t,\vect{l}) = & \sum_{\vect{x}} \epsilon^{abc} (C \gamma_5)_{\beta \gamma} \widetilde{u}^a_{\alpha,\vect{0}}(t,\vect{x}) \widetilde{u}^b_{\beta,\vect{0}}(t,\vect{x}) \widetilde{d}^c_{\gamma,\vect{l}}(t,\vect{x}) \,,
\end{align}
where $C$ is the charge conjugation matrix. The $\psi_{\Sigma^+}$ operator is the same with the $d$ quark exchanged with an $s$ quark. The smeared quark fields are defined by
\begin{align}
    \widetilde{q}_{\vect{l}}(t,\vect{x}) = \mathcal{N} \sum_{\vect{x}'} e^{-|\vect{x}-\vect{x}'|_L^2/2\sigma^2} e^{i \vect{l}\cdot \vect{x}} q(t,\vect{x}') \,,
\end{align}
where $\mathcal{N}$ is a normalisation factor, $\sigma$ is the smearing radius, and all Dirac and colour indices have been left implied as they are unaffected by the smearing procedure. The magnitude of lattice site separation is defined as the smallest distance between the 2 points taking into account the periodic boundary conditions $|\vect{x}-\vect{x}'|_L^2 = \sum_i (\min\{|x_i-x'_i|,L-|x_i-x'_i|\})^2$. For this work we use a smearing radius of $\sigma = 6a$ which was found to reduce overlap with excited states in the two-point functions.

\begin{figure}
    \centering
    \vspace{-0.7cm}
    \begin{multicols}{4}
        \null \includegraphics[width=\linewidth]{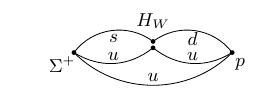} $C_{sd}$ \par
        \null \includegraphics[width=\linewidth]{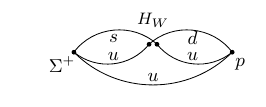} $C_{su}$ \par
        \includegraphics[width=\linewidth]{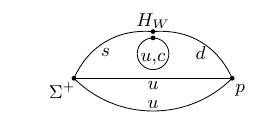} $E$ \par
        \null \vspace{-0.2cm} \includegraphics[width=\linewidth]{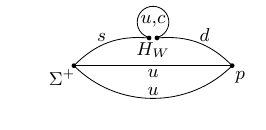} $S$ \par
    \end{multicols}
    \caption{Wick contraction topologies for the weak Hamiltonian three-point function. The contractions for the four-point function correspond to inserting an electromagnetic current on each leg, which are shown in \cref{ap:4pt_diagrams}.}
    \label{fig:Hw_topos}
\end{figure}

All measurements for this work were made using the Grid \cite{Boyle2016Grid} and Hadrons \cite{Portelli2020Hadrons} C++ libraries, where the contractions have been implemented.
The contractions relevant for the four-point function \eqref{eq:4pt_def} come in 4 distinct topologies (illustrated by the the weak Hamiltonian three-point function) shown in \cref{fig:Hw_topos}. The E and S diagrams, which we refer to as eye-type diagrams, require propagators that loop back to their starting position. The other two ($C_{sd}$ and $C_{su}$) are referred to as non-eye (NE) type diagrams. The four-point function contains these same topologies, but requires an additional electromagnetic current insertion on each of the quark legs, as well as a disconnected diagram, where the electromagnetic current couples to a sea quark. The full set of four-point function diagrams is shown in \cref{ap:4pt_diagrams}. In this calculation we neglect the fully disconnected electromagnetic current loop diagrams as they introduce additional noise, and are expected to be colour and $SU(3)$ flavour suppressed relative to the remaining diagrams. However, a precision calculation would require these diagrams be included.

We computed the quark loops in the eye diagrams using hypercubic sparsened $\mathbb{Z}_2 \otimes \mathbb{Z}_2$ noise sources, as are used in \cite{Boyle:2022ccj}, and we employ the All-Mode-Averaging (AMA) variance reduction approach \cite{Blum:2012uh} with 4 inexact hits per configuration, and a single exact hit per configuration for the bias correction. Due to the different statistical characteristics of the eye and non-eye type diagrams resulting from the stochastic loops, it is instructive to present quantities that utilise only the less noisy non-eye diagrams (denoted with a superscript NE), as well as using the full set of diagrams. The non-eye only quantities are not physical on their own, but can be interpreted as a quantity from a partially quenched theory with additional degenerate quark flavours in which certain Wick contractions are not present.
These non-eye quantities are instructive because their variance is dominated by gauge-noise which has a baryonic signal-to-noise ratio problem, while we find the eye diagrams in our calculation/setup to have a variance dominated by the stochastic $\mathbb{Z}_2$ noise of the loop.

The contraction method used for this calculation requires solving the quark propagators sourced at the baryon source and sink positions. In order to project to definite momentum with the gauge-fixed Gaussian sources used, this would require Dirac operator inversions at every spatial lattice site, which would be prohibitively expensive. Instead we take only lattice sites from a sparsened spatial lattice. This was introduced in \cite{Detmold:2019fbk} where the sparsened lattice was taken to be a coarser sublattice. In the latter study \cite{Li:2020hbj}, it was shown that a random sparsening of the lattice prevents contamination from additional momentum modes with only a small increase in statistical uncertainties. We therefore employ this random sparsening at the baryonic source and sink independently in order to obtain the best momentum projection and maximal statistics. A study of this technique in the context of this decay can be found in \cite{Hodgson:2021lzt}.

We compute these quark propagators on every other timeslice (i.e., on $32$ out of the total $T/a=64$), constructing four-point functions as defined in \cref{eq:4pt_def} that are then averaged to yield a single estimator per configuration. Similarly, we compute and average the $p$ and $\Sigma^+$ two-point functions from these sources to obtain one estimator per configuration. We compute the non-eye and eye type diagrams with a source-sink separation of ${\Delta t = t_p-t_\Sigma = 16a}$, and the non-eye diagrams with a second separation of $\Delta t = 12 a$. The electromagnetic current is always inserted at the midpoint $t_J - t_\Sigma \in \{ 6a, 8a \}$.

Our kinematics setup is chosen to maintain similarities with any future physical point calculation. As is described in \cite{Erben:2022tdu}, the finite-volume correction step is greatly simplified by the definite parity quantum number of the intermediate $N\pi$ states when the initial $\Sigma^+$ is at rest ($\vect{k}=\vect{0}$), which we choose to also use here even though such a finite-volume correction is not required at our unphysical pion mass. With the initial state momentum fixed, the final state momentum determines the momentum transfer $q^2$. Ideally we would have $q^2=0$ for a direct comparison with phenomenology. However, on the lattice, momentum values are restricted to discrete values unless twisted boundary conditions are employed. Additionally, accessing the form factors requires at least one of the hadrons to have non-zero momentum. By selecting a single unit of lattice momentum, $\vect{p}=\frac{2\pi}{L}(1,0,0)$ we achieve the closest possible value to $q^2=0$ on this lattice, yielding $q^2 = -0.207(2) \, \Gev^2$.

The Wilson coefficients in the weak Hamiltonian are reused from the work \cite{Christ:2012se}, where the known values in the $\overline{\text{MS}}$ renormalisation scheme \cite{Buchalla1995WeakLogarithms} are converted into the non-perturbative  RI/SMOM scheme,
\begin{align}
    C^{\textrm{lat}}(a) = C^{\overline{\textrm{MS}}}(\mu) Z^{\overline{\textrm{MS}} \to \textrm{lat}}(a,\mu) \, ,
\end{align}
where a mass scale of $\mu=2.15$ GeV is used throughout. This gives the Wilson coefficients for the bare four-quark operators on the lattice $C^\text{lat}_1 = -0.2216$ and $C^\text{lat}_2 = 0.6439$ \cite{Christ:2012se}.

\subsection{Form Factor Extraction}
The weak Hamiltonian \eqref{eq:weakH} contains a parity conserving and parity changing component
\begin{align}
    H_W = H_W^+ + H_W^- \,,
\end{align}
where $H_W^+$ contains four-quark operators with only the VV+AA gamma structures, while $H_W^-$ contains the VA+AV structures, for example $Q_1^q = Q_1^{q(+)} + Q_1^{q(-)}$ with
\begin{align}
    Q_1^{q(+)} & = [\bar{s} \gamma_\mu d] \, [\bar{q} \gamma^\mu q] + [\bar{s} \gamma_\mu \gamma_5 d] \, [\bar{q} \gamma^\mu \gamma_5 q] \\
    Q_1^{q(-)} & = -\left( [\bar{s} \gamma_\mu \gamma_5 d] \, [\bar{q} \gamma^\mu q] + [\bar{s} \gamma_\mu  d] \, [\bar{q} \gamma^\mu \gamma_5 q] \right) \,.
\end{align}
These two components can be computed separately, and each contain contributions from only two distinct form factors ($a,c$ from $H_W^+$ and $b,d$ from $H_W^-$). As is described in \cref{ap:negative_parity}, we observe no significant signal for both the non-eye and eye contributions to the $H_W^-$ four-point functions, and therefore we restrict ourselves to only the parity conserving sector from this point forward, extracting form factors $a, c$. The superscript $+$ on the weak Hamiltonian shall be left implied.

From the appendix of ref. \cite{Erben:2022tdu}, it can be seen that the form factors can be extracted using traces of the amplitude $\widetilde{\mathcal{A}}_\mu$ multiplied with suitable combinations of gamma matrices. The traces relevant for our kinematics have the form
\begin{align}
    \Tr[ \mathbb{P}_p \widetilde{\mathcal{A}}_\mu^X \mathbb{P}_\Sigma P^+ \gamma] = \zeta_{\mu,\gamma} f_\mu^X \,,
\end{align}
where $\gamma$ is a generic Dirac matrix and $P^+ = \frac{1}{2}(1+\gamma_t)$ is the positive parity projector. $\zeta_{\mu,\gamma}$ is a coefficient that absorbs kinematic factors related to the specific $\gamma$ used, and $f_\mu$ is a linear combination of the form factors $a$ and $c$. The superscript $X \in \{\rho,\sigma\}$ indicates the separation between the two time orderings contributing to the amplitude, and $f_\mu = f_\mu^\rho + f_\mu^\sigma$.

For this study with the $\Sigma^+$ at rest and the proton's momentum $\vec{p}=(p_x,0,0)$ along the $x$-axis, we choose to measure the amplitude components along the temporal and z axes, and use the gamma structures $\mathbbm{1}$ and $\gamma_y \gamma_5$ for these components respectively. The relevant coefficients are therefore
\begin{align}
    \zeta_{t,\mathbbm{1}} = - \frac{p_x^2}{m_p} \hspace{2em} \text{and} \hspace{2em} \zeta_{z,\gamma_y \gamma_5} = -i \frac{p_x}{m_p} \,,
\end{align}
which are equivalent to those listed in appendix C in \cite{Erben:2022tdu} up to an overall factor $4 m_\Sigma$ (accounted for by the projector normalisation) and the different sign since $\zeta_{z,\gamma_y \gamma_5} = -\zeta_{y,\gamma_z \gamma_5}$.
Numerical results presented in \cref{sec:results} will already have these traces taken and the coefficients $\zeta_{\mu,\gamma}$ removed.

Once the linear combinations of form factors $f_t$ and $f_z$ are obtained, the form factors can be extracted simply by inverting the linear system
\begin{align}
\label{eq:linear_relation_ff_ac}
    \left( \begin{array}{c}
        f_t \\
        f_z
    \end{array} \right) = \left( \begin{array}{cc}
        1 & m_\Sigma + m_p \\
        m_\Sigma + m_p & q^2
    \end{array} \right) \left( \begin{array}{c}
        a \\
        c
    \end{array} \right) \,.
\end{align}

\subsection{Fitting Methods}
\label{sec:Extraction_Methods}
Several methods to remove problematic intermediate-state exponentials and extract amplitudes or form factors are detailed in \cite{Erben:2022tdu}. In our case, with the unphysically large pion mass employed, there is only a single state that is exponentially growing: the single-proton intermediate state in $I_\mu^\rho$. However, due to the slowly decaying contribution from the single $\Sigma^+$ state in $I_\mu^\sigma$, this must also be accounted for to improve the convergence to the $T_b \to \infty$ limit. These exponentials can be removed using various techniques, such as explicitly constructing the exponentials from measurements of two- and three-point functions (as described in Eq.\ (3.59) of \cite{Erben:2022tdu}) or shifting the weak Hamiltonian by a scalar operator \cite{Christ2015ProspectsDecays,Erben:2022tdu}. However, these methods require additional measurements, which introduce their own complications and can lead to added statistical and/or systematic uncertainties. A discussion of these methods and their results are given in \cref{ap:alt_methods}.

To simplify the analysis, we adopt a more straightforward approach: including the intermediate-state exponential (corresponding to the integration variable $\omega$ in \cref{eq:Int_rho} or \cref{eq:Int_sig} being at the proton or $\Sigma^+$ mass respectively) directly in the fits to the integrated correlators. This approach, referred to as the direct fit method in \cite{Christ2016FirstKtopiell+ell-}, results in the following fit ansatz:
\begin{align}
\label{eq:fit_func}
    \Tr[I_\mu^X(T) P^+ \gamma]/\zeta_{\mu,\gamma} = f^X_\mu + \sum_n c^{X,(n)}_\mu \, e^{-(E_n-E^X) T} \,,
\end{align}
truncated over intermediate states, $n$, to only include the lowest in each spectrum ($p$ with momentum $\vect{k}$ in $I_\mu^\rho$ and $\Sigma^+$ with momentum $\vect{p}$ in $I_\mu^\sigma$). Here we define $E^\rho = E_\Sigma(\vect{k})$ and $E^\sigma = E_p(\vect{p})$.
This leaves the exponential coefficient, $c^{X,(n)}_\mu$, and energy difference, $E_n - E^X$, as fittable parameters along with $f^X_\mu$. However, these intermediate state parameters can be fixed using information about the energies and matrix elements. Fixing both is equivalent to the explicit construction method described in \cite{Erben:2022tdu}.
We instead fix the energy difference from measurements of the individual energies from two-point functions, while leaving only the $f^X_\mu$ and the $c_\mu^{X,(0)}$ as the fittable parameters.
It should be noted that in order to perform correlated fits to the integrated correlators, the covariance matrix must be estimated and inverted. However, if the full integrated correlator is used $I_\mu(T_a,T_b) = I_\mu^\rho(T_a) + I_\mu^\sigma(T_b)$, as was done in calculations of the rare kaon decay \cite{Christ2016FirstKtopiell+ell-,Boyle:2022ccj}, the relevant covariance matrix will be singular due to a redundancy in the data as is shown in \cref{ap:cov_mat}. This can be simply remedied by fitting instead to the spectrally separated integrated four-point functions defined in \cref{eq:Int_rho,eq:Int_sig}, as we do in this work.

%% file: Sections/Results.tex
\subsection{Two-point functions}
In order to extract the amplitude from the four-point function in \cref{eq:4pt_def}, the energies and creation/annihilation matrix elements of the external baryons are required. These are obtained from the trace of the two-point functions in \cref{eq:2pt_def} with a positive parity projector which has the ground state dominated form
\begin{align}
\label{eq:2pt_decomp}
    \Tr[\Gamma^{(2)}_B(t) P^+] = L^3 \left( 1+\frac{m_B}{E_B(\vect{l})} \right) |Z_B(\vect{l})|^2 e^{-E_B(\vect{l}) t} + O(e^{-E^\text{ex}t})
\end{align}
where $E^\text{ex}$ is the energy of the lightest excited state with the same quantum numbers as the baryon $B$.

For each baryon, a fully correlated simultaneous fit is performed to the two-point functions with momenta $\vect{l}=\vect{k}=\vect{0}$ and $\vect{p} =\frac{2\pi}{L}(1,0,0)$. The fit includes two types of correlators: those where both the source and sink quark fields are smeared, and those with a smeared source and a point sink. A ground state fit ansatz is used (that shown in \cref{eq:2pt_decomp} with excited states ignored), and we impose the continuum dispersion relation $E_B(\vect{l}) = \sqrt{m_B^2 + \vect{l}^2}$. This leaves 5 fit parameters $|Z_B(\vect{0})|$, $|Z_B(\vect{p})|$, $|Z_B^\text{pt}(\vect{0})|$, $|Z_B^\text{pt}(\vect{p})|$ and $m_B$.

\begin{figure}[h!]
    \centering
    \includegraphics{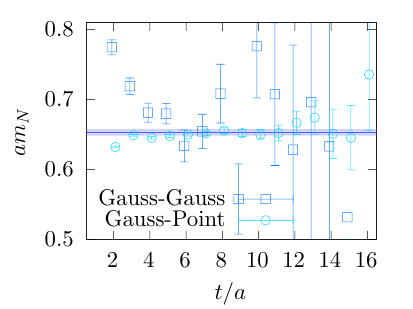}
    \includegraphics{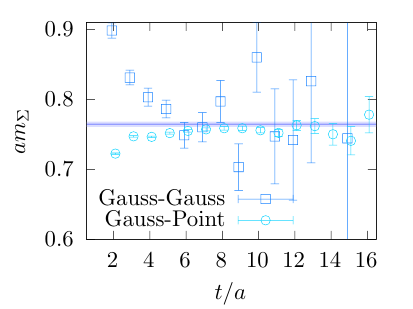}
    \caption{Effective masses of the stationary nucleon (left) and $\Sigma$ (right) baryons with gauge-fixed Gaussian smearing in the source, and with and without smearing in the sink. The horizontal band indicates the fit result of the corresponding mass. Note the fit is performed to the correlators and not directly to the effective masses shown. The $\chi^2/dof$ values for the fits are $27/20$ and $25/21$, corresponding to the left and right panel.
    }
    \label{fig:2pt_meff}
\end{figure}

Due to the use of a ground state ansatz, the fit must be performed in a region with no significant excited state effects. As can be seen in the effective mass plots in \cref{fig:2pt_meff}, for the smeared-smeared correlator this is satisfied from approximately time $t \gtrsim 6a$, while the smeared-point correlators reach a plateau within statistics later at approximately $t \gtrsim 10a$. We use this to choose the fit ranges that are given in \cref{tab:2pt_ranges}.
\begin{table}[]
    \centering
    \begin{tabular}{c|c|c|c}
        $B$ & Sink & $\vect{0}$ & $\vect{p}$ \\
        \hline
        \hline
        \multirow{2}{*}{$N$} & sm & $[6,15]$  & $[6,11]$ \\
         & pt & $[10,15]$ & $[10,14]$ \\
        \hline
        \multirow{2}{*}{$\Sigma$} & sm & $[6,14]$  & $[6,14]$ \\
         & pt & $[12,15]$ & $[12,15]$ \\
         \hline
    \end{tabular}
    \caption{Fit ranges $[t_{min},t_{max}]$ used for the two-point functions. The first column lists the baryon corresponding to the two-point function, while the second column indicates the sink type: either a smeared sink (``sm") or a point sink (``pt"). As explained in the text, the source is smeared in all two-point functions. The third column provides the fit ranges for zero-momentum two-point functions, and the fourth column lists those for the $\vect{p} =\frac{2\pi}{L}(1,0,0)$ two-point function.}
    \label{tab:2pt_ranges}
\end{table}

\subsection{Four-point functions}
\begin{figure}[h!]
    \centering
    \includegraphics{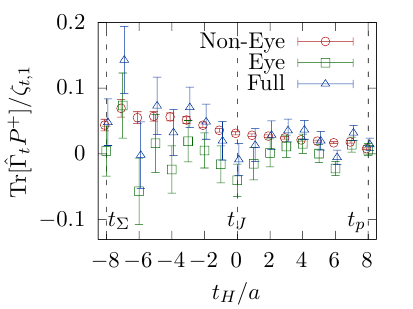}
    \includegraphics{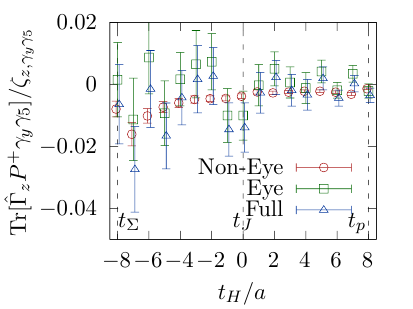}
    \caption{Temporal (left, with $\gamma_t$ Dirac structure in the electromagnetic current) and spatial (right, with $\gamma_z$ Dirac structure in the electromagnetic current) components of the traced positive parity $H_W^+$ four-point function, computed with a source-sink separation of $\Delta t/a=16$, plotted as a function of $t_H/a$. The individual contributions from the non-eye and eye diagrams are shown separately, along with their combined total, labelled as `Full'.}
    \label{fig:4pt}
\end{figure}
\begin{figure}[h!]
    \centering
    \includegraphics{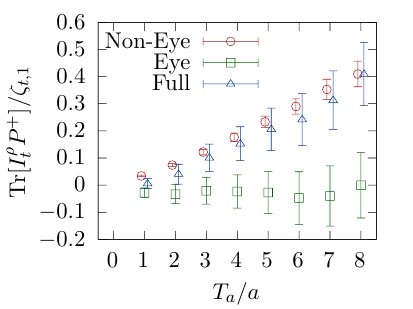}
    \includegraphics{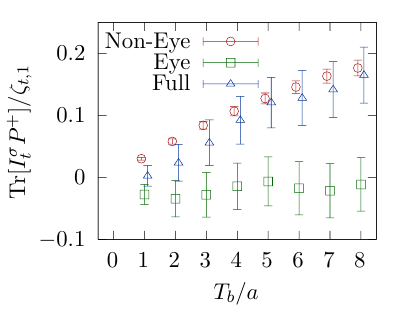}
    \includegraphics{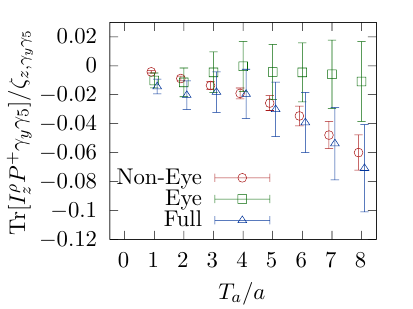}
    \includegraphics{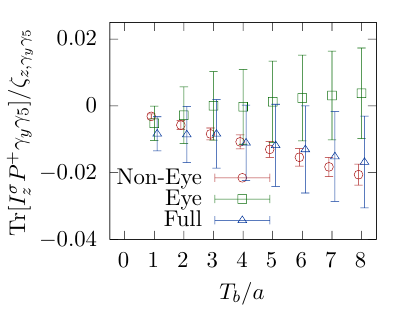}
    \caption{Temporal (top) and spatial (bottom) components of the traced integrated four-point functions in with a source-sink separation $\Delta t/a=16$. The left and right plots correspond to the two time orderings $I_\mu^\rho(T_a)$ and $I_\mu^\sigma(T_b)$ respectively. The horizontal axis represents the integration limits $T_a / a$ and $T_b / a$. Specifically, the left-hand plots show the cumulative sum of the datapoints to the left of $t_H/a = 0$ in \cref{fig:4pt}, while the right-hand plots show the cumulative sum of the datapoints to the right of this point. Contributions from the non-eye and eye diagrams are displayed separately, along with the total combined correlator.}
    \label{fig:4pt_Int}
\end{figure}

In \cref{fig:4pt}, we present the positive-parity $H_W^+$ part of the four-point function in \cref{eq:4pt_def} with a time separation of $\Delta t = 16a$, computed as outlined in the the previous section. Given the significantly different signal-to-noise characteristics of the two contributions, we separately display the eye diagrams (the combined $E$ and $S$ topologies from \cref{fig:Hw_topos}) and the non-eye diagrams (the combined $C_{sd}$ and $C_{su}$ topologies). Errors are estimated using the statistical bootstrap method.

The results show a clear signal for the non-eye diagrams, while the estimators for the eye diagrams are largely consistent with zero within $1 \sigma$ on most timeslices, and within  $2 \sigma$ on all timeslices. Since the signal from the non-eye diagrams is of a similar magnitude to the noise level in the eye diagrams, the total estimator for the four-point function - being the sum of both contributions - is dominated by noise and fluctuates around zero.

Fortuitously, the situation improves when we consider the integrated four-point function, as defined in \cref{eq:Int_rho,eq:Int_sig}. The individual contributions (both temporal and spatial, in both time orderings) are shown in \cref{fig:4pt_Int}. While the signal from the eye contribution remains consistent with zero, the central value of the less noisy non-eye contribution increases, benefiting from the summation over many precise datapoints in \cref{fig:4pt}. As a result, the central value of the more precise non-eye contribution surpasses the noise level of the eye diagrams, revealing a clear (albeit noisy) signal in the combined total of the two contributions. When summing the noisy datapoints of the raw four-point function to produce the integrated version, the noise adds in quadrature (up to correlations) which grows more slowly than the linear addition of the signal. This effect is more pronounced in the temporal component, where the signal is distinct, compared to the spatial component, where the signal is present but less clear.

This indicates that the full integrated-correlator may show a hierarchy between the non-eye and eye diagrams, which is very different behaviour to what was observed in the calculations of the rare kaon decay $K^+ \to \pi^+ \ell^+ \ell^-$ \cite{Christ2016FirstKtopiell+ell-,Boyle:2022ccj}, where the non-eye and eye contributions have a similar magnitude and opposite sign leading to a large cancellation.
The presence of such a hierarchy in this decay process suggests that the non-eye diagrams alone may serve as a reasonable approximation of the total. It is important to emphasize that such an approximation is only empirically justified and the systematic effects of neglecting the eye diagrams remains uncontrolled. Nonetheless, we will use it in the following analysis as a tool for obtaining a qualitative estimate of the true result. We emphasize that even if this approximation proves to be incorrect, and the eye diagrams are not a sub-dominant contribution, focusing on the signal from the NE diagrams alone remains valuable. It provides an indication of the potential precision of the results if an effective noise reduction method for the eye diagrams were identified. Efforts to mitigate noise in such loop diagrams, as introduced in~\cite{Giusti:2019kff}, are currently being explored in the context of rare kaon decays~\cite{Lattice24:hodgson}.

\subsection{Form Factors}
\begin{figure}[h!]
    \centering
    \includegraphics{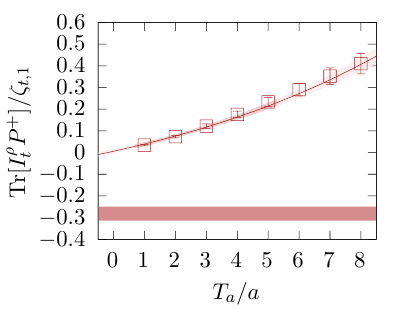}
    \includegraphics{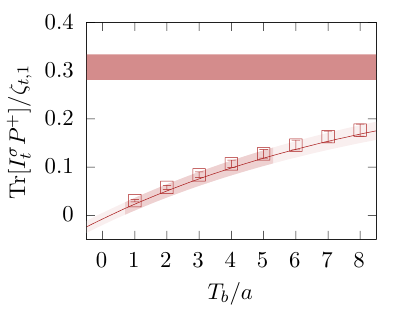}
    \includegraphics{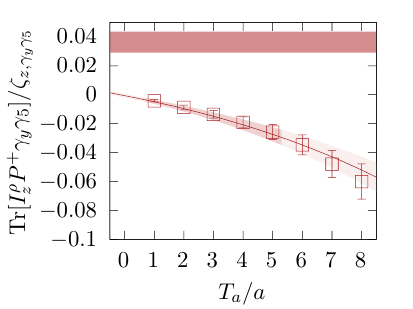}
    \includegraphics{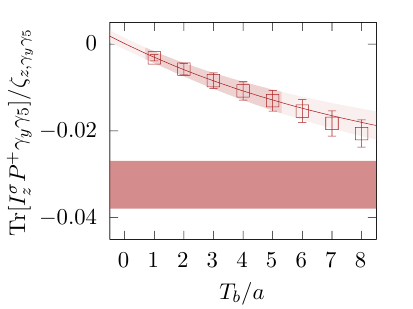}
    \caption{Temporal (top) and spatial (bottom) components of the non-eye contribution to the integrated four-point function with source-sink separation $\Delta t/a = 16$. Datapoints are identical to the red points in \cref{fig:4pt_Int}, and, as in that figure, the left and right plots correspond to the two time orderings $I_\mu^\rho$ and $I_\mu^\sigma$, respectively. Overlaid are fits to this data using the fit form \cref{eq:fit_func} for a single state $n$. The darker shaded areas of the fit reconstructions represent the range within which the data being fit lies. The horizontal band represents the extracted $f^X_\mu$, where $X \in \{\rho, \sigma\}$. The $\chi^2/dof$ values for the fits are $6.0/3, 2.8/3, 1.6/3$ and $6.3/2$, corresponding to the four panels in standard reading order: top left, top right, bottom left, and bottom right.
    }
    \label{fig:4pt_Int_fit_NE}
\end{figure}

\begin{figure}[h!]
    \centering
    \includegraphics{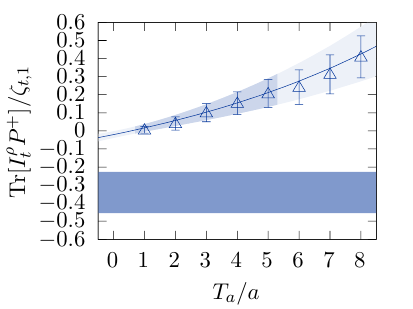}
    \includegraphics{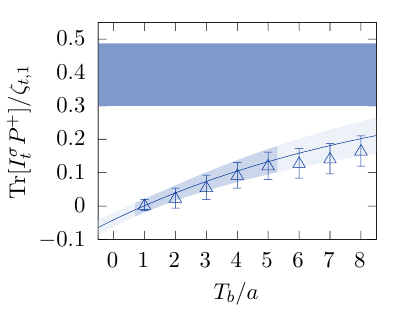}
    \includegraphics{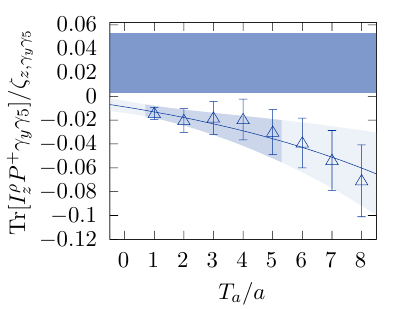}
    \includegraphics{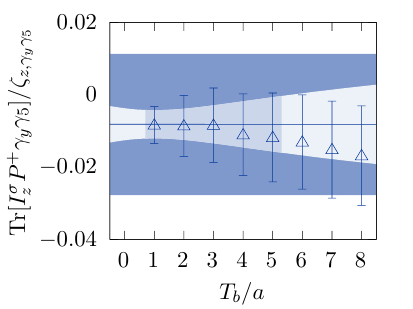}
    \caption{Similar to \cref{fig:4pt_Int_fit_NE}, but using the datapoints corresponding to the total integrated four-point function (i.e., the blue datapoints in \cref{fig:4pt_Int}) with overlaid fits to the data. The $\chi^2/dof$ values for the fits are $2.3/3, 1.2/3, 1.9/3$ and $1.5/3$, corresponding to the four panels in standard reading order: top left, top right, bottom left, and bottom right.
    }
    \label{fig:4pt_Int_fit_Tot}
\end{figure}

We extract the linear combinations $f_\mu^X$ of the form factors $a$ and $c$, where $X \in \{\rho, \sigma\}$, from fits to the integrated correlators with functional form as in \cref{eq:fit_func} for a single state $n$. We show these $f_\mu^X$ as well as fit reconstructions overlaying the data in \cref{fig:4pt_Int_fit_NE} for the non-eye contribution as well as in \cref{fig:4pt_Int_fit_Tot} for the total contribution. The numerical values of $f_\mu^X$, extracted from these fits, are shown in \cref{tab:results_f}.
\begin{table}
    \centering
        \begin{tabular}{c|c|c|c}
            & $f^\rho_t \,[10^{\fitexp{RH_NE_rho_muT}}]$ & $f^\sigma_t \,[10^{\fitexp{RH_NE_sig_muT}}]$ & $f_t = f_t^\rho + f_t^\sigma  \,[10^{\fitexp{RH_NE_muT}}]$ \\
            \hline
            \hline
            NE   & $\fitvalerrNoexp{RH_NE_rho_muT}$ & $\fitvalerrNoexp{RH_NE_sig_muT}$ & $\fitvalerrNoexp{RH_NE_muT}$ \\
            \hline
            Full   & $\fitvalerrNoexp{RH_Tot_rho_muT}$ & $\fitvalerrNoexp{RH_Tot_sig_muT}$ & $\fitvalerrNoexp{RH_Tot_muT}$ \\
            \hline
            \multicolumn{4}{c}{} \\
            & $f^\rho_z \,[10^{\fitexp{RH_NE_rho_muZ}}]$ & $f^\sigma_z \,[10^{\fitexp{RH_NE_sig_muZ}}]$ & $f_z = f_z^\rho + f_z^\sigma \,[10^{\fitexp{RH_NE_muZ}}]$ \\
            \hline
            \hline
            NE   & $\fitvalerrNoexp{RH_NE_rho_muZ}$ & $\fitvalerrNoexp{RH_NE_sig_muZ}$ & $\fitvalerrNoexp{RH_NE_muZ}$ \\
            \hline
            Full   & $\fitvalerrNoexp{RH_Tot_rho_muZ}$ & $\fitvalerrNoexp{RH_Tot_sig_muZ}$ & $\fitvalerrNoexp{RH_Tot_muZ}$ \\
            \hline
        \end{tabular}
    \caption{Fit results for linear combinations of form factors $f_\mu^X$, presented for both the non-eye contribution only (`NE') and the full contribution (`Full'). These values correspond to the fit results shown in \cref{fig:4pt_Int_fit_NE,fig:4pt_Int_fit_Tot}, where they are represented as horizontal bands. Also included is the sum of both time orderings, $f_\mu$, which exhibits a significant cancellation between $f^\rho_\mu$ and $f^\sigma_\mu$. All four-point functions used in the calculation were computed with a source-sink separation of $\Delta t/a = 16$.}
    \label{tab:results_f}
\end{table}

As can be inferred from both the plots and the table, we obtain a clear signal for many of the spectrally separated quantities measured. For the temporal form factor $f_t^X$, the individual components for $X \in \{\rho, \sigma\}$ have an error of approximately $10\%$ for the non-eye diagrams only and $30\%$ for the full contribution, which includes both the eye and non-eye diagrams. The spatial component is noisier, with a signal at around $20\%$ for the non-eye only part and a result (almost) consistent with zero for the full contribution. However, the central values of $f_\mu^\rho$ and $f_\mu^\sigma$ in all cases are of similar magnitude but opposite sign, leading to a significant cancellation in the relevant sum of form factors $f_\mu = f_\mu^\rho + f_\mu^\sigma$, which results in a signal compatible with zero, even for the non-eye diagrams alone, both in the temporal and spatial components. Whether this cancellation is caused by some underlying mechanism or is accidental is currently unknown and worth exploring in future work. Further investigation may lead to alternative methods for directly computing the sum, avoiding the statistical issues associated with such cancellations.

Another important caveat is that our assumption regarding the suppression of the eye diagrams relative to the non-eye diagrams was based on the individual time orderings,  $f_\mu^\rho$ and $f_\mu^\sigma$, rather than their sum. Since the individual components exhibit a cancellation, this assumption may be less robust. Due to the large statistical uncertainties, it remains uncertain whether a similar cancellation would occur in the eye diagram contributions to the combined form factors $f_\mu$.

The positive parity rare hyperon form factors $a$ and $c$ can be obtained from $f_t$ and $f_z$ by inverting \cref{eq:linear_relation_ff_ac}. In our case, the form factors $a$ and $c$, and hence their linear combinations $f_\mu$, are real because intermediate $N\pi$ states are absent under the unphysical masses used. Consequently, only the real parts of $a$ and $c$ are obtained and are given in \cref{tab:ffs}. Since all $f_\mu$ are compatible with zero within their quoted errors, the resulting values of $a$ and $c$ are also consistent with zero.

Despite this, we can draw some comparisons to available phenomenological calculations~\cite{He2005DecayModel,He2018DecayPmu+mu-,Roy:2024hqg}. It is important to emphasize that while these phenomenological values are obtained for the physical pion mass and $q^2 = 0$, our results are derived at a heavier pion mass of 340 MeV, a finite momentum transfer of $q^2 = -0.2 \, \mathrm{GeV}^2$, and with a single lattice spacing. As such, our results are still subject to quark-mass and discretization effects. Nevertheless, the values obtained from the non-eye diagrams (see \cref{tab:ffs}) are of the same order of magnitude as the phenomenological ones:
\begin{align}
    \Re a \sim 10 \, \Mev \hspace{3em} \text{and} \hspace{3em} \Re c \sim 10^{-2} \, .
\end{align}
Under the (uncontrolled) assumption that the non-eye diagrams dominate the central value, as suggested by the integrated four-point functions, the proximity to the phenomenological values indicates that the non-eye diagrams alone may be close to resolving a signal in this unphysical setup.

Improving the non-eye diagrams alone will not address the dominant error from the eye diagrams. In principle, quark loops could be calculated with higher statistics using additional noise hits, but this approach is costly due to the large error, which scales as $\sqrt{N}$. A more efficient strategy would involve improved stochastic estimators for loop propagators. For example, the split-even estimator described in~\cite{Giusti:2019kff} directly computes the difference between light and charm loop propagators and has shown significant variance reduction in observables with quark loops. This method is under investigation for the very similar rare kaon decay $K \to \pi \ell^+ \ell^-$~\cite{Lattice24:hodgson} and could potentially reduce the variance of the eye diagrams for rare hyperon decays to subdominant levels in future studies.

Even once stochastic noise is addressed, rare hyperon decay calculations remain signal-limited by gauge noise. To explore how close this calculation is to resolving a signal in the absence of stochastic noise, the non-eye results can serve as a proxy. Given the exponential degradation of the signal-to-noise ratio for baryonic quantities, reducing the source-sink separation could enhance the signal.

\begin{table}
    \centering
    \begin{tabular}{c|c|c}
         & $\Re a \, [\Mev]$ & $\Re c \, [10^{\fitexp{RH_NE_c}}]$ \\
        \hline
        \hline
        NE & $\fitvalerrNoexp{RH_NE_a}$ & $\fitvalerrNoexp{RH_NE_c}$ \\
        \hline
        Full & $\fitvalerrNoexp{RH_Tot_a}$ & $\fitvalerrNoexp{RH_Tot_c}$ \\
        \hline
    \end{tabular}
    \caption{Results for the real part of the parity-conserving form factors $a$ and $c$, calculated using only the non-eye diagrams, as well as the full set of diagrams. These are derived from the linear combinations $f_\mu$ of the form factors listed in \cref{tab:ffs} using the relation given in \cref{eq:linear_relation_ff_ac}. All four-point functions used in the calculation were computed with a source-sink separation of $\Delta t/a = 16$.}
    \label{tab:ffs}
\end{table}

\begin{figure}
    \centering
    \includegraphics{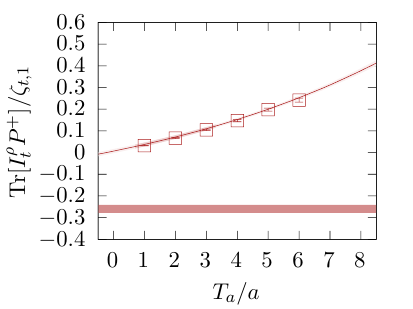}
    \includegraphics{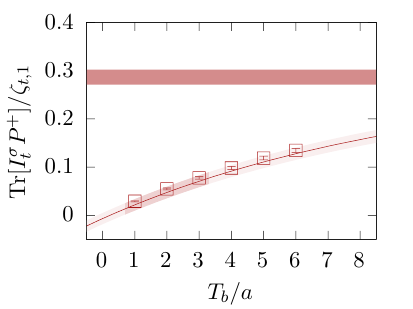}
    \includegraphics{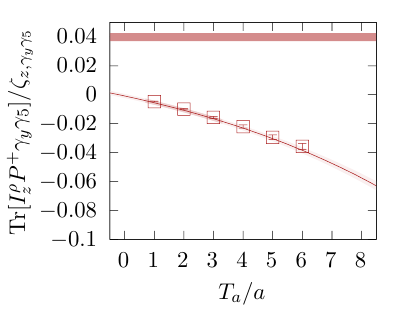}
    \includegraphics{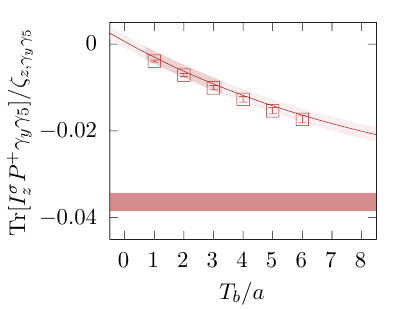}
    \caption{Similar to \cref{fig:4pt_Int_fit_NE}, but using a four-point function with a baryon separation of $\Delta t = 12a$. The $\chi^2/dof$ values for the fits are $0.8/1, 4/1, 1.1/1$ and $0.02/1$, corresponding to the four panels in standard reading order: top left, top right, bottom left, and bottom right.
    }
    \label{fig:4pt_Int_fit_dt12}
\end{figure}

\Cref{tab:ffs_alt} presents results for the form factors computed using only the non-eye diagrams, with a reduced baryon separation of $\Delta t/a = 12$. The data and fit reconstruction are shown in \cref{fig:4pt_Int_fit_dt12}. Due to decreased gauge noise, a signal at the $40\%-50\%$ level can be observed, even after cancellation between the time orderings. However, this improved statistical precision introduces an increased and challenging-to-quantify systematic error in the form of excited-state contamination.
There is extensive discussion in the literature on the issue of handling excited state effects in nucleon matrix elements, an overview of which is given in Section 10 of \cite{FLAG:2024oxs}. Several approaches have been introduced to approach this problem including multi-state fits \cite{RQCD:2019jai,He:2021yvm,Gupta:2021ahb,Agadjanov:2023efe,Jang:2023zts,Alexandrou:2024ozj}, the summation method \cite{MAIANI1987420,GUSKEN1989266,Alexandrou:2024ozj} and closely related the Feynman-Hellman method \cite{CSSM:2014uyt,Bouchard:2016heu,Savage:2016kon,QCDSFUKQCDCSSM:2023qlx}, and GEVP improved interpolators \cite{Barca:2022uhi,Alexandrou:2024tin,Barca:2024hrl}. Note these reference lists are by no means exhaustive.

All of these methods require additional data in the form of many source-sink separations or multiple initial/final state interpolators, which in this project would massively inflate the cost, and therefore we attempt to gain some basic insight into the excited state effects from the two-point functions. While it is well understood that the dominant excited states between two- and higher-point correlation functions can be very different due to enhanced matrix elements \cite{Tiburzi:2009zp,Bar:2016uoj,Hansen:2016qoz,Bar:2017kxh,Bar:2017gqh,Bar:2018xyi,Jang:2019vkm,Bar:2019gfx,Bar:2021crj}, we can at least get an approximate idea of the timescales at which excited states dominate from the two-point function effective masses shown in \cref{fig:2pt_meff}, in which the ground state dominance appears at source-sink separations of $t \gtrsim 6a$ with smeared sources and sinks. This suggests that for the external baryon ground states to dominate, the $H_W$ and $J_\mu$ operators should be separated from the baryon interpolators by at least 6 timeslices. Clearly the results shown in \cref{tab:ffs_alt} obtained with $\Delta t/a = 12$ leave no room to meet this condition since the vector current is located halfway between them at $t_J-t_\Sigma = t_p - t_J = 6a$, and the weak Hamiltonian will always be closer than that to either one of the external baryons depending on time ordering. It is therefore clear that at $\Delta t/a = 12$, the resulting form factors most likely have significant excited state contamination.

\begin{table}
    \centering
    \begin{tabular}{c|c|c}
         & $\Re a \, [\Mev]$ & $\Re c \, [10^{\fitexp{RH_NE_c}}]$ \\
        \hline
        \hline
        NE $\Delta t/a=12$ & $\fitvalerrNoexp{RH_NE_dt12_a}$ & $\fitvalerrNoexp{RH_NE_dt12_c}$ \\
        \hline
    \end{tabular}
    \caption{Similar to \cref{tab:ffs}, but using four-point functions computed with a source-sink separation of $\Delta t/a = 12$. This results in more precise values for $\Re a$ and $\Re c$, leading to the emergence of a signal in the non-eye diagrams.}
    \label{tab:ffs_alt}
\end{table}

This excited state contamination is reduced, but likely still large, at the larger source-sink separation of $\Delta t/a = 16$. Therefore future calculations of the rare hyperon decay must increase the source-sink separation to mitigate this systematic effect, while simultaneously reducing gauge noise to resolve a signal. Addressing these challenges simultaneously will be difficult and/or costly. While advancements in computing technology may help reduce costs, future calculations would greatly benefit from algorithmic improvements, particularly those targeting exponential signal-to-noise challenges. One promising direction is the use of multi-level algorithms~\cite{Luscher:2001up}, though significant development would be required before they could be applied to the rare hyperon decay.

%% file: Sections/Summary.tex
In summary, we have presented the results of the first lattice calculation of positive parity form factors of the rare hyperon decay $\Sigma^+ \to p \ell^+ \ell^-$, utilising the methods of \cite{Erben:2022tdu}. This calculation has been performed away from the physical point with a pion mass of $m_\pi = 340 \, \Mev$ using domain-wall fermions. While the calculation is highly challenging and the results are affected by significant statistical uncertainties, this work serves as an important proof-of-principle. Notably, our results for the non-eye contribution are of the same order of magnitude as phenomenological estimates. Additionally, we observe empirical evidence suggesting that the more statistically noisy eye contribution may be suppressed in magnitude, which is very different to the behaviour observed in the rare kaon decay \cite{Christ2016FirstKtopiell+ell-,Boyle:2022ccj}.

By far the largest uncertainty comes from the stochastic estimation of loop propagators within the eye type diagrams. Simply computing additional statistics would be a highly inefficient approach to overcoming these uncertainties, and therefore investigation into improved methods of computing quark loops is required. One promising method that is already showing large improvements for different observables is the split-even estimator for quark loop difference \cite{Giusti:2019kff}. There is evidence that this method is effective for estimating the amplitude of the rare kaon decay $K \to \pi \ell^+ \ell^-$~\cite{Lattice24:hodgson}. While this decay is closely related to the present case of the rare hyperon decay $\Sigma^+ \to p \ell^+ \ell^-$, its effectiveness in improving results for this specific observable has yet to be demonstrated.

The next most dominant uncertainty is the statistical gauge noise of all diagrams. Due to the baryonic nature of the correlation functions in this calculation, there exists an exponential signal-to-noise ratio problem. In addition to this well understood source of noise, we find a large cancellation between the two time orderings contributing to the amplitude which each have a well resolved contribution (when restricted to only the non-eye diagrams). This cancellation reduces this relatively good signal on the separated parts, to no significant signal on the total. If the cause of this cancellation can be identified, it may be possible to construct a new methodology that avoids such cancellation and to compute the total directly. 

Due to the baryonic signal-to-noise problem, larger source-sink separations observe significantly worse signal, and therefore we also analyse a shorter source-sink separation with improved signal to show that our calculation is not far from overcoming this subdominant statistical error. This does however come at the cost of increasing the systematic error coming from excited state contamination, which is likely already rather large. 
It is therefore clear that in order to achieve a significant result, even at the unphysical point, all of these challenges must be overcome, which will take considerable effort. 

Looking further ahead, a physical point calculation will suffer more greatly from these problems due to the lighter pion mass and heavier charm mass, and therefore scalable solutions to these challenges must be found. In addition, at the physical point the $N\pi$ intermediate states are below threshold and therefore a calculation requires treatment of $N\pi$ scattering and the $\Sigma^+ \to N\pi$ decay in order to account for the power-like finite volume effects and additional growing intermediate exponentials.

Due to these challenges, a physical lattice computation of the rare hyperon decay is unlikely within in the foreseeable future, however, efforts to overcome the many challenges in this process will likely be of great benefit to the lattice community for a wide array of other observables.

One of the key motivations for pursuing this decay using lattice QCD was the significant challenges faced by phenomenological approaches in obtaining precise estimates for the rare hyperon decay form factors. While, as discussed above, achieving a precision result from lattice QCD may remain out of reach with current state-of-the-art methodology and computational resources, this complementary approach still holds promise for making meaningful contributions. In particular, if the application of noise reduction techniques to rare hyperon decays proves successful, it could significantly mitigate the noise in the eye diagrams. This advancement would enable the determination of the sign of $\Re a$, reducing an ambiguity present in the current best phenomenological approaches. In combination with an experimental measurement of the polarised $\Sigma \to p \gamma$ decay, a lattice determination of the sign of $\Re a$ could uniquely resolve the 4-fold ambiguity currently limiting the branching fraction. These developments would mark an important step forward in our understanding of rare hyperon decays and demonstrate the potential of lattice QCD in tackling similarly complex processes.

%% file: Sections/Ap_Cov_Mat.tex
The separation of the integrated four-point function into two parts each containing only a single time ordering ($t_H \lessgtr 0$) given in \cref{eq:Int_rho,eq:Int_sig} not only gives us access to each operator ordering of the amplitude separately, it also has significant implications for the numerical evaluation of the amplitude. For this appendix we leave the subscript $\mu$ implicit on the integrals as it is irrelevant for this discussion.

The full integral in \cref{eq:Int_all} is a function of two time indices $I(T_a,T_b)$, and therefore the covariance matrix is indexed by two pairs of indices
\begin{align}
    \cov(I,I)_{(i_a,i_b),(j_a,j_b)} \,,
\end{align}
where $i_a,j_a \in \{1,\dots,N_a\}$ and $i_b,j_b \in \{1,\dots,N_b\}$ index the $N_a$ and $N_b$ values of $T_a$ and $T_b$ respectively.
It is also helpful to define the covariance between the separated integrated four-point functions as a block matrix
\begin{align}
    \cov (I^\rho|I^\sigma) = \left( \begin{array}{c|c} \text{cov}(I^\rho,I^\rho) & \text{cov}(I^\rho,I^\sigma) \\
    \hline
    \text{cov}(I^\sigma,I^\rho) & \text{cov}(I^\sigma,I^\sigma) \end{array} \right) \,,
\end{align}
that is indexed by $k \in \{ 1,\dots,N_a,N_a+1,\dots,N_a+N_b\}$ where the first $N_a$ elements correspond to the index space of $I^\rho(T_a)$ while the last $N_b$ correspond to the index space of $I^\sigma(T_b)$.
Since $I(T_a,T_b) = I^\rho(T_a) + I^\sigma(T_b)$, the covariance matrix of the full integral can be expanded out as
\begin{align}
    & \cov(I,I)_{(i_a,i_b),(j_a,j_b)} \\
    & = \cov(I^\rho,I^\rho)_{i_a,j_a} + \cov(I^\rho,I^\sigma)_{i_a,j_b} + \cov(I^\sigma,I^\rho)_{i_b,j_a} + \cov(I^\sigma,I^\sigma)_{i_b,j_b} \\
    & = \cov(I^\rho|I^\sigma)_{i_a,j_a} + \cov(I^\rho|I^\sigma)_{i_a,j_b+N_a} + \cov(I^\rho|I^\sigma )_{i_b+N_a,j_a} + \cov(I^\rho|I^\sigma)_{i_b+N_a,j_b+N_a} \,,
\end{align}
and therefore can be written in matrix notation as
\begin{align}
\label{eq:Cov_relation}
    \text{cov}(I,I) = P \, \cov(I^\rho|I^\sigma) \, P^T \,,
\end{align}
where $P$ is the rectangular matrix
\begin{align}
    P_{(i_a,i_b),k} = (\delta_{i_a,k}+\delta_{i_b+N_a,k}) \,.
\end{align}
It is clear that $\cov(I,I)$ has dimension $N_aN_b$, while $\cov(I^\rho|I^\sigma)$ has dimension $N_a+N_b$, and therefore
\begin{align}
    \text{rank}(\cov(I,I)) \leq \text{rank}(\cov(I^\rho|I^\sigma)) \leq N_a+N_b \,.
\end{align}
So long as $N_a+N_b < N_a N_b$, which is the case for all practical instances, then $\text{rank}(\cov(I,I)) < \text{dim}(\cov(I,I))$ which guarantees that $\det(\cov(I,I)) = 0$ and therefore the covariance matrix of the full integral is uninvertible. Since the inverse correlation matrix is required for performing correlated fits, fits to the full integrated correlator must be partially or fully decorrelated. This is due to a redundancy of information contained within $I(T_a,T_b)$ that is removed when separated into its constituents $I^\rho(T_a)$ and $I^\sigma(T_b)$.

In the lattice calculation of the rare kaon decay \cite{Boyle:2022ccj}, 2-dimensional fits were performed to $I(T_a,T_b)$, and it was observed that performing correlated fits was not possible, which is a manifestation of this singular correlation matrix. In this work we fit to the two separated integrals $I^\rho(T_a)$ and $I^\sigma(T_b)$ which allows for correlated fits to be performed.

%% file: Sections/Ap_Discrete_Time.tex
The formulae provided in \cite{Erben:2022tdu} assume continuous time, however, in practice lattice calculations must be performed at a finite lattice spacing with discrete time, and therefore the integral over the time variable must be replaced with some discretised sum over points separated by the lattice spacing $a$.
To demonstrate the different discretisation approaches, we consider the discretised integral of a simple exponential
\begin{align}
    I(T) = \int_0^T dt \, e^{-\omega t} = \frac{1-e^{-\omega T}}{\omega} \,,
\end{align}
that can then be related to all of the formulae regarding the integrated four-point correlation function. The simplest discretisation is the Riemann sum
\begin{align}
    I_a^{(1)}(T>0) = a \sum_{n=0}^{T/a-1} e^{-\omega a n} = a \frac{1-e^{-\omega T}}{1-e^{-\omega a}} \,.
\end{align}
Removing the $T$ dependent part, as is done to extract the amplitude, we find the kernel that the spectral functions are convolved with is given by
\begin{align}
    K_a^{(1)}(\omega) = \frac{a}{1-e^{-\omega a}} = \frac{1}{\omega} (1+O(a \omega)) \,,
\end{align}
which has cut-off effects of $O(a \omega)$ relative to the continuum kernel $1/\omega$. In addition, since we split the full integral into two, recombining them with this discretisation double counts the point $t_H = 0$, which can be seen as a cut-off effect.

An improvement to this discretisation is to instead use the trapezium rule which takes the form
\begin{align}
    I_a^{(2)}(T>0) = a \left( \frac12 e^{- \omega \cdot 0} + \sum_{n=1}^{T/a-1} e^{-\omega an} + \frac12 e^{-\omega T} \right) = a \frac{1+e^{-\omega a}}{2} \frac{1-e^{-\omega T}}{1-e^{-\omega a}} \,.
\end{align}
The resultant kernel is given by
\begin{align}
    K_a^{(2)}(\omega) = \frac{a}{2} \frac{1+e^{-\omega a}}{1-e^{-\omega a}} = \frac{1}{\omega} (1+ O(a\omega)^2) \,,
\end{align}
which has improved cut-off effects. Again in our separated integral setup, this can be viewed as sharing the point $t_H = t_J = 0$ equally between the two integrals. We choose to use this discretisation for this work, however, one could in principle use a higher order integrator such as Simpson's rule (as is done in \cite{Gerardin:2023naa}) to decrease these effects further
\begin{align}
    I_a^{(4)}(T>0) = \frac{a}{3} \left( e^{-\omega \cdot 0} + 4 \sum_{n=1,\text{odd}}^{T/a-1} e^{-\omega an} + 2 \sum_{n=2,\text{even}}^{T/a-2} e^{-\omega an} + e^{-\omega T} \right) \,,
\end{align}
which corresponds to the kernel
\begin{align}
    K_a^{(4)}(\omega) = \frac{a}{3} \frac{1+4 e^{-\omega a} + e^{-2 \omega a}}{1- e^{-2 \omega a}} = \frac{1}{\omega}(1+O(a\omega)^4) \,.
\end{align}
Improving the integration discretisation reduces the cut-off effects induced by the amplitude extraction methods, but not those associated with the energies and matrix elements in the four-point function. Additionally, improving the integration generally reduces the number of fittable points available (e.g. only even values of $T/a$ can be used with the Simpson integration), and it is therefore a trade-off that must be made for a given observable.

\begin{figure}
    \centering
    \includegraphics[width=0.6\linewidth]{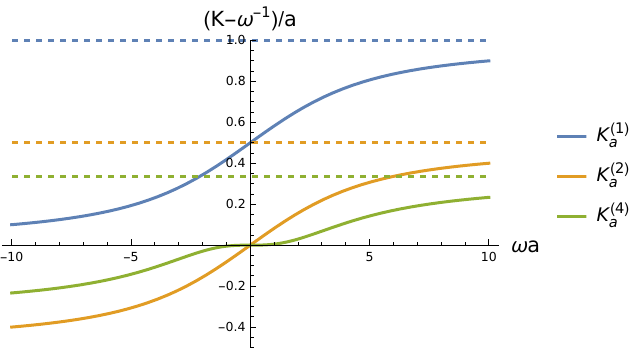}
    \caption{Cut-off effects on the integration kernel for the Riemann sum ($K^{(1)}_a$), Trapezium rule ($K^{(2)}_a$), and Simpson's rule ($K^{(4)}_a$) in units of the inverse lattice spacing. The dashed lines indicate the limiting value as $\omega \to \infty$.}
    \label{fig:Int_kern_err}
\end{figure}

Finally, it should be noted that the order of the cut-off effects has been quoted in powers of $a \omega$. However, since $\omega$ gets integrated over in this case, these powers will get arbitrarily large while the true error relative to the continuum kernel plateaus to a constant for $|\omega| \gg a^{-1}$. It is therefore instructive to view the exact cut-off effects of the kernel over a range of $\omega$, which is shown in \cref{fig:Int_kern_err}.

%% file: Sections/AP_fourptdiagrams.tex
\begin{figure}[h]
\centering
\includegraphics[width=0.65\textwidth]{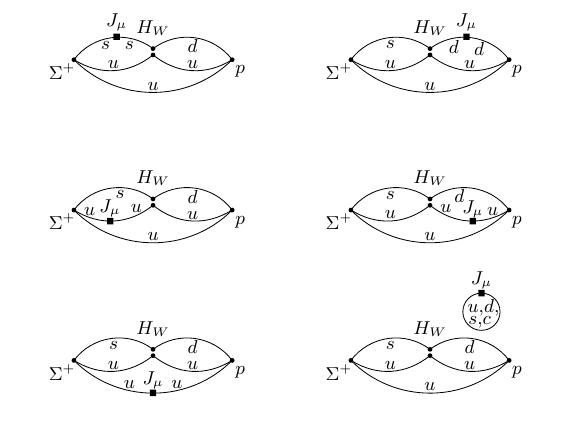}
\caption{Four-point functions for $C_{sd}$ topology.}
\label{fig:4pt_Csd_topo}
\end{figure}
\begin{figure}[h]
\centering
\includegraphics[width=0.65\textwidth]{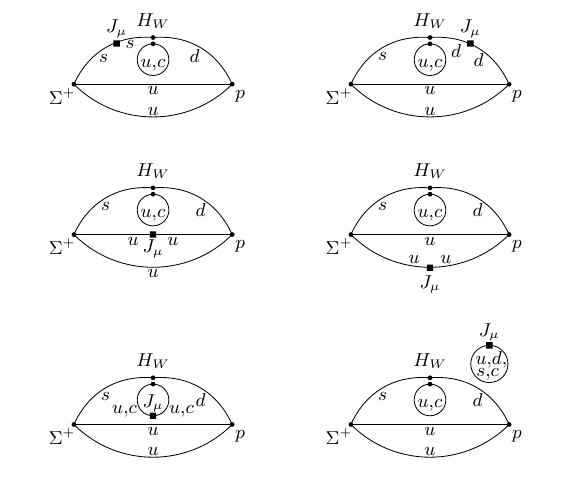}
\caption{Four-point functions for $E$ topology.}
\label{fig:4pt_E_topo}
\end{figure}
\begin{figure}[h]
\centering
\includegraphics[width=0.65\textwidth]{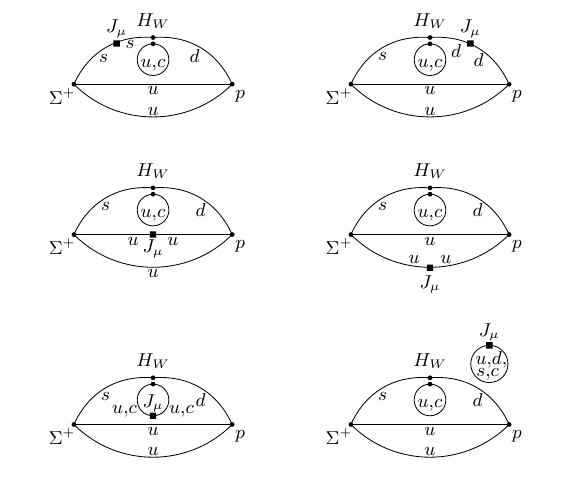}
\caption{Four-point functions for $S$ topology.}
\label{fig:4pt_S_topo}
\end{figure}

There are 24 Wick contraction entering into the computation of the four-point correlation function in \cref{eq:4pt_def}. These can be separated into the non-eye type ($C_{sd}$ and $C_{su}$ topologies in \cref{fig:4pt_Csd_topo,fig:4pt_Csd_topo}) and the eye type ($E$ and $S$ in \cref{fig:4pt_E_topo,fig:4pt_S_topo}).

\begin{figure}
\centering
\includegraphics[width=0.65\textwidth]{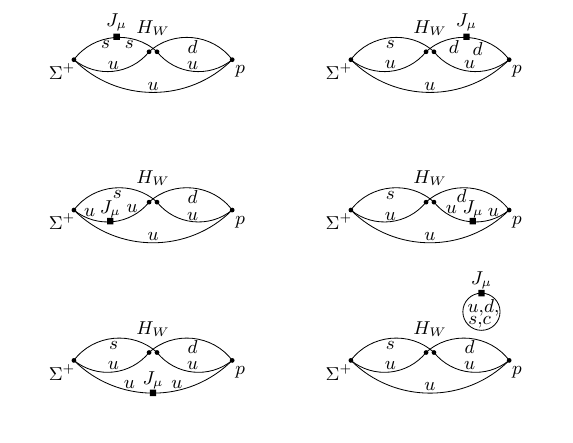}
\caption{Four-point functions for $C_{su}$ topology.}
\label{fig:4pt_Csu_topo}
\end{figure}

%% file: Sections/Ap_Negative_Parity.tex
\begin{figure}
    \centering
    \includegraphics{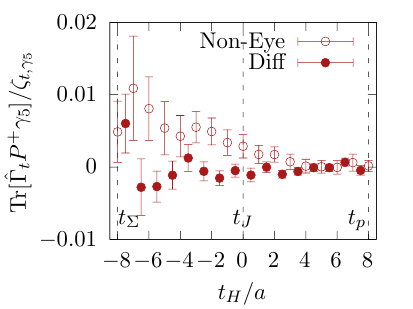}
    \includegraphics{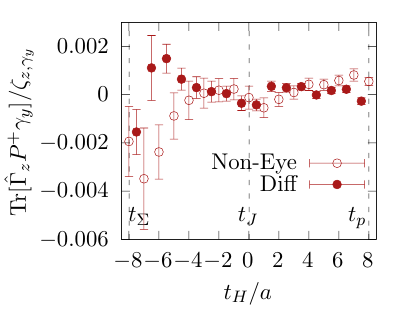}
    \caption{Temporal (left) and spatial (right) components of the non-eye part of the traced negative parity four-point function in \cref{eq:4pt_def}, computed with a source-sink separation of $\Delta t/a=16$, plotted as a function of $t_H/a$.}
    \label{fig:4pt_negative_parity}
\end{figure}

\Cref{fig:4pt_negative_parity} shows the non-eye contribution to the four-point correlation functions in the negative parity channel (i.e. using the $H_W^-$ operator), in which no clear signal is observed. This is in contrast with the similar observation in the positive parity channel shown in \cref{fig:4pt} in which these non-eye diagrams show a clear signal. There does appear to be some structure within \cref{fig:4pt_negative_parity}, however, the correlated differences between neighbouring times are consistent with zero, suggesting that this structure is consistent with correlated statistical fluctuations.
Therefore, the negative-parity form factors cannot be reliably extracted from these data, and the addition of the eye-type diagrams only compounds this issue to the point that no meaningful information can be extracted in this channel from the existing data.
We have therefore decided to only analyse the positive-parity channel in this work.

%% file: Sections/Ap_Alt_Methods.tex
The results quoted for this project in \cref{sec:results} utilise the direct fit method to remove the time dependence of the intermediate states in order to access the amplitude. However, there exist other potential methods described in \cite{Erben:2022tdu}. Here we investigate these alternative methods and discuss their challenges. Due to the degradation of signal coming from the eye-type diagrams, we investigate these methods using only the much more precise non-eye diagrams.

\subsection{Explicit Construction}
The Explicit Construction method works by fixing the energy difference and $c^{X,(n)}_\mu$ coefficients in the fit form \cref{eq:fit_func} from measurements of two- and three-point functions. As is the case in the main text, we restrict ourselves to only accounting for the lightest intermediate state for each time-ordering. By examining the matrix elements that enter into these coefficients, it can be seen that
\begin{align}
    c_\mu^{\rho,(p)} & = \frac{1}{\zeta_{\mu,\gamma}} \frac{m_p}{E_p(\vect{k})}\frac{\Tr[ \mathbb{P}_p(\vect{p}) \widetilde{\mathcal{J}}_\mu^{p}(\vect{q}) \mathbb{P}_p(\vect{k}) [h(\vect{k})] \mathbb{P}_\Sigma(\vect{k}) P^+ \gamma]}{E_p(\vect{k})-E_\Sigma(\vect{k})} \\
    c_\mu^{\sigma,(\Sigma^+)} & = \frac{1}{\zeta_{\mu,\gamma}} \frac{m_\Sigma}{E_\Sigma(\vect{p})}\frac{\Tr[ \mathbb{P}_p(\vect{p}) [h(\vect{p})] \mathbb{P}_\Sigma(\vect{p}) \widetilde{\mathcal{J}}_\mu^{\Sigma}(\vect{q}) \mathbb{P}_\Sigma(\vect{k}) P^+ \gamma]}{E_\Sigma(\vect{p})-E_p(\vect{p})} \,,
\end{align}
where $h$ and $\widetilde{\mathcal{J}}_\mu$ are related to the Weak Hamiltonian and electromagnetic current matrix elements by
\begin{align}
    \bra{p(\vect{l}),r} H_W \ket{\Sigma^+(\vect{l}),s} = & \bar{u}_p^r(\vect{l}) h(\vect{l}) u_\Sigma^s(\vect{l}) \\
    \bra{p(\vect{p}),r} J_\mu \ket{p(\vect{k}),s} = & \bar{u}_p^r(\vect{p}) \widetilde{\mathcal{J}}^p_\mu(\vect{q}) u_p^s(\vect{k}) \\
    \bra{\Sigma^+(\vect{p}),r} J_\mu \ket{\Sigma^+(\vect{k}),s} = & \bar{u}_\Sigma^r(\vect{p}) \widetilde{\mathcal{J}}^\Sigma_\mu(\vect{q}) u_\Sigma^s(\vect{k}) \,.
\end{align}

It is possible to measure these the $c_\mu^{X,(n)}$ objects via traces of products of three-point functions as is shown in \cite{Erben:2022tdu}. However, some assumptions allow us to separate this into the measurement of two separate three-point functions which significantly simplifies the analysis. These can also allow us to replace some measurements with more statistically precise measurements.

$h(\vect{l})$ is the weak Hamiltonian scalar form factor which is a function of the four-momentum difference squared $(k_\Sigma -k_p)^2=(E_\Sigma(\vect{l}) - E_p(\vect{l}))^2$, and therefore only depends on the three-momentum. Since the energy difference only takes values in a small $195 \, \text{MeV}$ range $0<E_\Sigma -E_p< m_\Sigma - m_p$, we can make the approximation that the form factor is independent of energy, which is exact in the $\text{SU}(3)$ flavour symmetric limit. This is the baryonic equivalent of the same approximation made in the rare kaon decay \cite{Christ2016FirstKtopiell+ell-}.
Using this approximation, we replace this form factor in $c_\mu^{\sigma,(\Sigma^+)}$ with the measurement at zero momentum where the best signal is observed. In principle it would be easy to remove this assumption and measure the form factor with non-zero momentum, however at our level of statistics this is not possible.

The second assumption made is that the product of different particle projectors can be written as a single projector
\begin{align}
    \mathbb{P}_p(\vect{p}) \mathbb{P}_\Sigma(\vect{p}) \simeq \mathbb{P}_p(\vect{p}) \,.
\end{align}
By their definition, it can be seen that this is exactly true in the $\vect{p}=\vect{0}$ limit, and that the corrections are $O(|\vect{p}|/m_\Sigma)$ for non-zero momentum.

Using both of these approximations, the $c_\mu^{\rho,(p)}$ and $c_\mu^{\sigma,(\Sigma)}$ coefficients become
\begin{align}
    c_\mu^p & \simeq \frac{1}{\zeta_{\mu,\gamma}} \frac{m_p}{E_p(\vect{k})}\frac{h(\vect{k}) \Tr[ \mathbb{P}_p(\vect{p}) \widetilde{\mathcal{J}}_\mu^{p} \mathbb{P}_p(\vect{k}) P^+ \gamma]}{E_p(\vect{k})-E_\Sigma(\vect{k})} \\
    c_\mu^\Sigma & \simeq \frac{1}{\zeta_{\mu,\gamma}} \frac{m_\Sigma}{E_\Sigma(\vect{p})}\frac{h(\vect{k}) \Tr[ \mathbb{P}_\Sigma(\vect{p}) \widetilde{\mathcal{J}}_\mu^{\Sigma} \mathbb{P}_\Sigma(\vect{k}) P^+ \gamma]}{E_\Sigma(\vect{p})-E_p(\vect{p})} \,,
\end{align}
where $h(\vect{k})$ and $\Tr[ \mathbb{P}_B(\vect{p}) \widetilde{\mathcal{J}}_\mu^{B} \mathbb{P}_B(\vect{k}) P^+ \gamma]$ can separately be measured from independent three-point functions
\begin{align}
    \Gamma^{(3)}_{H_W}(t_p,t_\Sigma) = & \langle \psi_p(t_p,\vect{k}) \, H_W(0) \, \overline{\psi}_\Sigma(t_\Sigma,\vect{k}) \rangle \\
    \Gamma^{(3)}_{\mu,B}(t',t) = & \langle \psi_B(t',\vect{p}) \, J_\mu(0) \, \overline{\psi}_B(t,\vect{k}) \rangle \,.
\end{align}

\subsection{Scalar Shift}
A second alternative method that can be used to remove a single intermediate state is the scalar shift method. This allows the positive and negative parity components of the Weak Hamiltonian to be shifted by a scalar or pseudoscalar operator respectively
\begin{align}
    H_W'^+ = H_W^+ - c_S \mathcal{S} \hspace{1em},\hspace{1em} H_W'^- = H_W^- - c_P \mathcal{P} \,,
\end{align}
where $\mathcal{S} = \bar{d}s$ and $\mathcal{P} = \bar{d} \gamma_5 s$, and $c_S$ and $c_P$ are arbitrary coefficients that can be chosen to eliminate the effects of a particular intermediate state. Since we restrict ourselves to the positive parity sector, only the scalar shift is relevant.

To remove the single proton intermediate state, the value of $c_S$ is chosen to be
\begin{align}
    c_S = \frac{ \bra{p(\vect{k}),r} H_W \ket{\Sigma^+(\vect{k}),s} }{ \bra{p(\vect{k}),r} \mathcal{S} \ket{\Sigma^+(\vect{k}),s} } \,,
\end{align}
which is measured from the ratio of three-point functions with the Weak Hamiltonian and scalar operators.

With this shift applied we use the same fit function as the direct fit method, however, the cancellation of the growing intermediate state should give a suppressed exponential coefficient. Due to the approximate $SU(3)_F$ symmetry of the octet baryons, this should also significantly suppress the single $\Sigma$ intermediate state in $I_\mu^\sigma$.

\subsection{Results}
\begin{figure}[h!]
    \centering
    \includegraphics{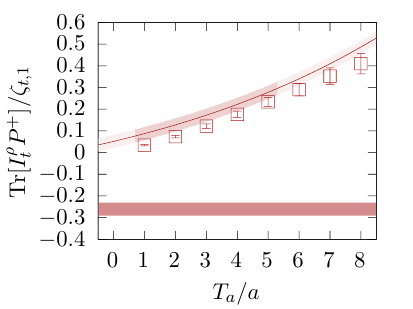}
    \includegraphics{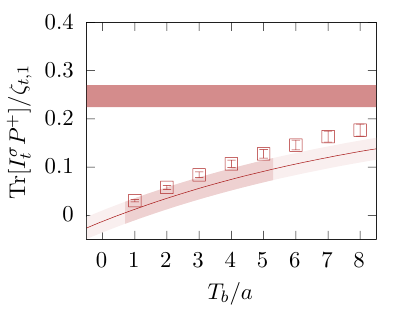}
    \includegraphics{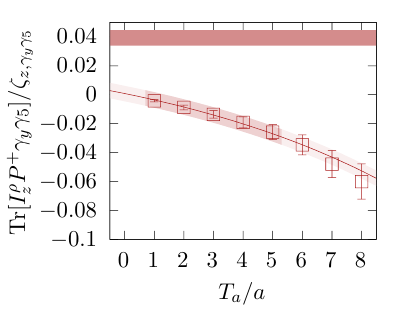}
    \includegraphics{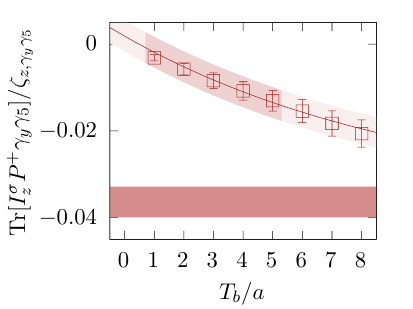}
    \caption{Similar to \cref{fig:4pt_Int_fit_NE}, but fit using the explicit construction method. The $\chi^2/dof$ values for the fits are $6.8/4, 10/4, 1.8/4$ and $6.4/4$, corresponding to the four panels in standard reading order: top left, top right, bottom left, and bottom right.
    }
    \label{fig:4pt_Int_fit_expl_rec}
\end{figure}

\begin{figure}[h!]
    \centering
    \includegraphics{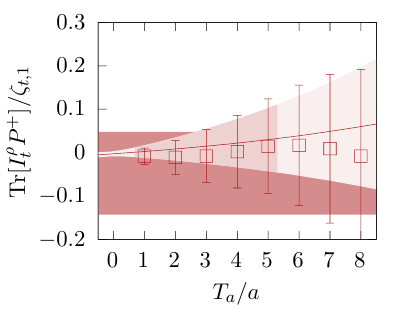}
    \includegraphics{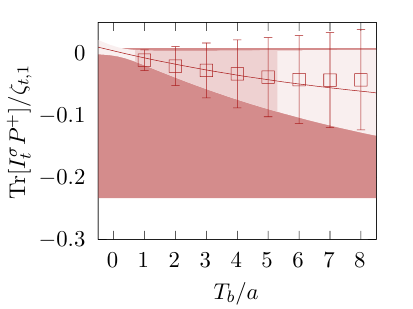}
    \includegraphics{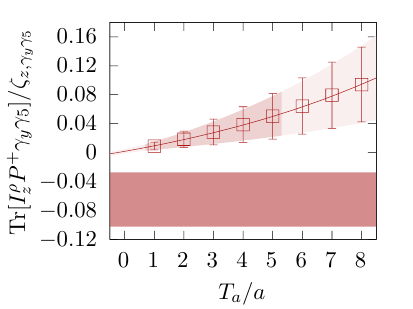}
    \includegraphics{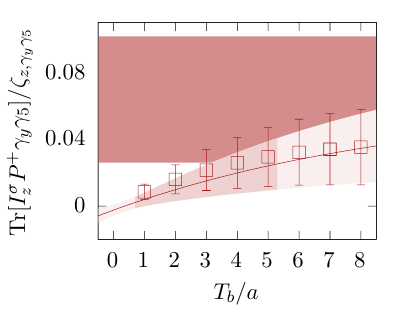}
    \caption{Similar to \cref{fig:4pt_Int_fit_NE}, but using a four-point function shifted by the scalar shift method. The $\chi^2/dof$ values for the fits are $2.2/3, 0.6/3, 0.8/3$ and $2.6/3$, corresponding to the four panels in standard reading order: top left, top right, bottom left, and bottom right.
    }
    \label{fig:4pt_Int_fit_scal_sub}
\end{figure}

We present the non-eye four-point function fits using the explicit construction method in \cref{fig:4pt_Int_fit_expl_rec} and fits to the the modified scalar shifted four-point function in \cref{fig:4pt_Int_fit_scal_sub}. Both figures follow the same layout as \cref{fig:4pt_Int_fit_NE} in the main text, which displays the unmodified non-eye four-point function. Additionally, they include reconstructions of the data based on the respective fit results. In \cref{tab:4pt_Int_fit_exp_construct}, we show the numerical values of the real parts of the form factors $a$ and $c$ obtained from the fits for $\Delta t=16$, alongside the direct fit non-eye $\Delta t=12,16$ results from the main text for reference.

Within the large uncertainties, all results appear mutually compatible, though there is a potential slight tension in the extraction of $\Re c$ using the scalar subtraction method compared to the other two approaches. The scalar shift method also results in the least constrained form factors, with error estimates more than four times larger than those obtained using our preferred method, where the single-proton state is explicitly included in the fit function. Explicitly reconstructing the intermediate states from three-point function data reduces errors slightly compared to our preferred method, but at the cost of including additional approximations stated above, and additional fitting systematics from the three-point functions.

The central values of both $\Re a$ and $\Re c$ shift between our preferred method and the explicit reconstruction method, with $\Re c$ even changing sign. However, within statistical errors, the results remain consistent. This indicates that even at $\Delta t=16$, our extraction method is subject to systematic effects, which are expected to be more pronounced at $\Delta t=12$, as discussed in the main text.

\begin{table}[h!]
    \centering
    \begin{tabular}{c|c|c|c}
         & $\Delta t/a$ & $\Re a \, [\Mev]$ & $\Re c \, [10^{\fitexp{RH_NE_c}}]$ \\
        \hline
        \hline
        NE Direct Fit & $12$ & $\fitvalerrNoexp{RH_NE_dt12_a}$ & $\fitvalerrNoexp{RH_NE_dt12_c}$ \\
        \hline
        NE Direct Fit & $16$ & $\fitvalerrNoexp{RH_NE_a}$ & $\fitvalerrNoexp{RH_NE_c}$ \\
        \hline
        NE Explicit Construct & $16$ & $\fitvalerrNoexp{RH_NE_meas_a}$ & $\fitvalerrNoexp{RH_NE_meas_c}$ \\
        \hline
        NE Scalar Subtract & $16$ & $\fitvalerrNoexp{RH_NE_sub_a}$ & $\fitvalerrNoexp{RH_NE_sub_c}$ \\
        \hline
    \end{tabular}
    \caption{Results for the real parts of form factors $a$ and $c$ obtained from non-eye four-point functions and using various extraction methods, as described in this appendix and in Section~\ref{sec:Extraction_Methods} of the main text. The first two rows present our preferred results obtained using the direct fit method. As in the main text, we report these results for both $\Delta t/a = 12$ and $\Delta t/a = 16$. The third row shows the result obtained by explicitly constructing the growing state, while the fourth row provides the result using the scalar shift method. Both of these approaches are detailed in this appendix, and their results are given only for $\Delta t/a = 16$. }
    \label{tab:4pt_Int_fit_exp_construct}
\end{table}

%% file: main.bbl
\providecommand{\href}[2]{#2}\begingroup\raggedright\begin{thebibliography}{10}

\bibitem{HyperCPCollaboration2005Evidencemu}
{\scshape HyperCP} collaboration, \emph{{Evidence for the decay $\Sigma^+ \to p \mu^+ \mu^-$}}, \href{https://doi.org/10.1103/PhysRevLett.94.021801}{\emph{Phys. Rev. Lett.} {\bfseries 94} (2005) 021801} [\href{https://arxiv.org/abs/hep-ex/0501014}{{\ttfamily hep-ex/0501014}}].

\bibitem{LHCb:2017rdd}
{\scshape LHCb} collaboration, \emph{{Evidence for the rare decay $\Sigma^+ \to p \mu^+ \mu^-$}}, \href{https://doi.org/10.1103/PhysRevLett.120.221803}{\emph{Phys. Rev. Lett.} {\bfseries 120} (2018) 221803} [\href{https://arxiv.org/abs/1712.08606}{{\ttfamily 1712.08606}}].

\bibitem{ParticleDataGroup:2024cfk}
{\scshape Particle Data Group} collaboration, \emph{{Review of particle physics}}, \href{https://doi.org/10.1103/PhysRevD.110.030001}{\emph{Phys. Rev. D} {\bfseries 110} (2024) 030001}.

\bibitem{LHCb-CONF-2024-002}
{\scshape LHCb} collaboration, \emph{{Observation of the $\Sigma^+ \to p \mu^+ \mu^-$ rare decay at LHCb}},  Tech. Rep. \href{https://cds.cern.ch/record/2899907}{LHCb-CONF-2024-002, CERN-LHCb-CONF-2024-002}, CERN, Geneva (2024).

\bibitem{Dettori:2297352}
F.~Dettori, D.~Martinez~Santos and J.~Prisciandaro, \emph{{Low-$p_T$ dimuon triggers at LHCb in Run 2}},  Tech. Rep. \href{http://cds.cern.ch/record/2297352}{LHCb-PUB-2017-023, CERN-LHCb-PUB-2017-023}, CERN, Geneva (2017).

\bibitem{LHCb:2025evf}
{\scshape LHCb} collaboration, \emph{{Observation of the very rare $\Sigma^+ \to p \mu^+ \mu^-$ decay}},  \href{https://arxiv.org/abs/2504.06096}{{\ttfamily 2504.06096}}.

\bibitem{He2005DecayModel}
X.-G.~He, J.~Tandean and G.~Valencia, \emph{{The Decay $\Sigma^+ \to p l^+ l^-$ within the standard model}}, \href{https://doi.org/10.1103/PhysRevD.72.074003}{\emph{Phys. Rev. D} {\bfseries 72} (2005) 074003} [\href{https://arxiv.org/abs/hep-ph/0506067}{{\ttfamily hep-ph/0506067}}].

\bibitem{He2018DecayPmu+mu-}
X.-G.~He, J.~Tandean and G.~Valencia, \emph{{Decay rate and asymmetries of $\Sigma^+\to p\mu^+\mu^-$}}, \href{https://doi.org/10.1007/JHEP10(2018)040}{\emph{JHEP} {\bfseries 10} (2018) 040} [\href{https://arxiv.org/abs/1806.08350}{{\ttfamily 1806.08350}}].

\bibitem{Roy:2024hqg}
A.~Roy, J.~Tandean and G.~Valencia, \emph{{\ensuremath{\Sigma^+}\textrightarrow{}\ensuremath{p\ell^+ \ell^-} decays within the standard model and beyond}}, \href{https://doi.org/10.1103/PhysRevD.111.013003}{\emph{Phys. Rev. D} {\bfseries 111} (2025) 013003} [\href{https://arxiv.org/abs/2404.15268}{{\ttfamily 2404.15268}}].

\bibitem{BESIII:2023fhs}
{\scshape BESIII} collaboration, \emph{{Precision Measurement of the Decay \ensuremath{\Sigma^+}\textrightarrow{}\ensuremath{p\gamma} in the Process J/\ensuremath{\psi}\textrightarrow{}\ensuremath{\Sigma^+\Sigma^-}}}, \href{https://doi.org/10.1103/PhysRevLett.130.211901}{\emph{Phys. Rev. Lett.} {\bfseries 130} (2023) 211901} [\href{https://arxiv.org/abs/2302.13568}{{\ttfamily 2302.13568}}].

\bibitem{BESIII:2023sgt}
{\scshape BESIII} collaboration, \emph{{Test of $C\!P$ Symmetry in Hyperon to Neutron Decays}}, \href{https://doi.org/10.1103/PhysRevLett.131.191802}{\emph{Phys. Rev. Lett.} {\bfseries 131} (2023) 191802} [\href{https://arxiv.org/abs/2304.14655}{{\ttfamily 2304.14655}}].

\bibitem{Shi:2025xkp}
R.-X.~Shi, Z.~Jia, L.-S.~Geng, H.~Peng, Q.~Zhao and X.~Zhou, \emph{{Status and prospect of weak radiative hyperon decays}},  \href{https://arxiv.org/abs/2502.15473}{{\ttfamily 2502.15473}}.

\bibitem{Erben:2022tdu}
F.~Erben, V.~G\"ulpers, M.T.~Hansen, R.~Hodgson and A.~Portelli, \emph{{Prospects for a lattice calculation of the rare decay $\Sigma^+ \to p \ell^+ \ell^-$}}, \href{https://doi.org/10.1007/JHEP04(2023)108}{\emph{JHEP} {\bfseries 04} (2023) 108} [\href{https://arxiv.org/abs/2209.15460}{{\ttfamily 2209.15460}}].

\bibitem{Christ2015ProspectsDecays}
{\scshape RBC/UKQCD} collaboration, \emph{{Prospects for a lattice computation of rare kaon decay amplitudes: $K\to\pi\ell^+\ell^-$ decays}}, \href{https://doi.org/10.1103/PhysRevD.92.094512}{\emph{Phys. Rev. D} {\bfseries 92} (2015) 094512} [\href{https://arxiv.org/abs/1507.03094}{{\ttfamily 1507.03094}}].

\bibitem{Briceno:2019opb}
R.A.~Brice\~no, Z.~Davoudi, M.T.~Hansen, M.R.~Schindler and A.~Baroni, \emph{{Long-range electroweak amplitudes of single hadrons from Euclidean finite-volume correlation functions}}, \href{https://doi.org/10.1103/PhysRevD.101.014509}{\emph{Phys. Rev. D} {\bfseries 101} (2020) 014509} [\href{https://arxiv.org/abs/1911.04036}{{\ttfamily 1911.04036}}].

\bibitem{Buchalla1995WeakLogarithms}
G.~Buchalla, A.J.~Buras and M.E.~Lautenbacher, \emph{{Weak decays beyond leading logarithms}}, \href{https://doi.org/10.1103/RevModPhys.68.1125}{\emph{Rev. Mod. Phys.} {\bfseries 68} (1996) 1125} [\href{https://arxiv.org/abs/hep-ph/9512380}{{\ttfamily hep-ph/9512380}}].

\bibitem{Christ2016FirstKtopiell+ell-}
N.H.~Christ, X.~Feng, A.~J{\"u}ttner, A.~Lawson, A.~Portelli and C.T.~Sachrajda, \emph{{First exploratory calculation of the long-distance contributions to the rare kaon decays $K\to\pi\ell^+\ell^-$}}, \href{https://doi.org/10.1103/PhysRevD.94.114516}{\emph{Phys. Rev. D} {\bfseries 94} (2016) 114516} [\href{https://arxiv.org/abs/1608.07585}{{\ttfamily 1608.07585}}].

\bibitem{Boyle:2022ccj}
P.A.~Boyle, F.~Erben, J.M.~Flynn, V.~G\"ulpers, R.C.~Hill, R.~Hodgson et~al., \emph{{Simulating rare kaon decays $K^{+}\to\pi^{+}\ell^{+}\ell^{-}$ using domain wall lattice QCD with physical light quark masses}},  \href{https://arxiv.org/abs/2202.08795}{{\ttfamily 2202.08795}}.

\bibitem{RBC:2010qam}
{\scshape RBC/UKQCD} collaboration, \emph{{Continuum Limit Physics from 2+1 Flavor Domain Wall QCD}}, \href{https://doi.org/10.1103/PhysRevD.83.074508}{\emph{Phys. Rev. D} {\bfseries 83} (2011) 074508} [\href{https://arxiv.org/abs/1011.0892}{{\ttfamily 1011.0892}}].

\bibitem{RBC:2012cbl}
{\scshape RBC/UKQCD} collaboration, \emph{{Domain Wall QCD with Near-Physical Pions}}, \href{https://doi.org/10.1103/PhysRevD.87.094514}{\emph{Phys. Rev. D} {\bfseries 87} (2013) 094514} [\href{https://arxiv.org/abs/1208.4412}{{\ttfamily 1208.4412}}].

\bibitem{RBC:2014ntl}
{\scshape RBC/UKQCD} collaboration, \emph{{Domain wall QCD with physical quark masses}}, \href{https://doi.org/10.1103/PhysRevD.93.074505}{\emph{Phys. Rev. D} {\bfseries 93} (2016) 074505} [\href{https://arxiv.org/abs/1411.7017}{{\ttfamily 1411.7017}}].

\bibitem{Boyle2016Grid}
P.A.~Boyle, G.~Cossu, A.~Yamaguchi and A.~Portelli, \emph{{Grid: A next generation data parallel C++ QCD library}}, \href{https://doi.org/10.22323/1.251.0023}{\emph{PoS} {\bfseries LATTICE2015} (2016) 023}.

\bibitem{Portelli2020Hadrons}
A.~Portelli, N.~Lachini, F.~Erben, M.~Marshall, F.~Joswig, R.~Hodgson et~al., \emph{aportelli/hadrons: Hadrons v1.4}, \href{https://doi.org/10.5281/zenodo.4063666}{\emph{Zenodo} (2023) }.

\bibitem{Blum:2012uh}
T.~Blum, T.~Izubuchi and E.~Shintani, \emph{{New class of variance-reduction techniques using lattice symmetries}}, \href{https://doi.org/10.1103/PhysRevD.88.094503}{\emph{Phys. Rev. D} {\bfseries 88} (2013) 094503} [\href{https://arxiv.org/abs/1208.4349}{{\ttfamily 1208.4349}}].

\bibitem{Detmold:2019fbk}
W.~Detmold, D.J.~Murphy, A.V.~Pochinsky, M.J.~Savage, P.E.~Shanahan and M.L.~Wagman, \emph{{Sparsening algorithm for multihadron lattice QCD correlation functions}}, \href{https://doi.org/10.1103/PhysRevD.104.034502}{\emph{Phys. Rev. D} {\bfseries 104} (2021) 034502} [\href{https://arxiv.org/abs/1908.07050}{{\ttfamily 1908.07050}}].

\bibitem{Li:2020hbj}
Y.~Li, S.-C.~Xia, X.~Feng, L.-C.~Jin and C.~Liu, \emph{{Field sparsening for the construction of the correlation functions in lattice QCD}}, \href{https://doi.org/10.1103/PhysRevD.103.014514}{\emph{Phys. Rev. D} {\bfseries 103} (2021) 014514} [\href{https://arxiv.org/abs/2009.01029}{{\ttfamily 2009.01029}}].

\bibitem{Hodgson:2021lzt}
R.~Hodgson, F.~Erben, V.~G\"ulpers and A.~Portelli, \emph{{Towards a lattice determination of the form factors of the rare Hyperon decay $\Sigma^+ \to p \ell^+ \ell^-$}}, \href{https://doi.org/10.22323/1.396.0480}{\emph{PoS} {\bfseries LATTICE2021} (2022) 480} [\href{https://arxiv.org/abs/2112.09599}{{\ttfamily 2112.09599}}].

\bibitem{Christ:2012se}
{\scshape RBC/UKQCD} collaboration, \emph{{Long distance contribution to the $K_L-K_S$ mass difference}}, \href{https://doi.org/10.1103/PhysRevD.88.014508}{\emph{Phys. Rev. D} {\bfseries 88} (2013) 014508} [\href{https://arxiv.org/abs/1212.5931}{{\ttfamily 1212.5931}}].

\bibitem{Giusti:2019kff}
L.~Giusti, T.~Harris, A.~Nada and S.~Schaefer, \emph{{Frequency-splitting estimators of single-propagator traces}}, \href{https://doi.org/10.1140/epjc/s10052-019-7049-0}{\emph{Eur. Phys. J. C} {\bfseries 79} (2019) 586} [\href{https://arxiv.org/abs/1903.10447}{{\ttfamily 1903.10447}}].

\bibitem{Lattice24:hodgson}
R.~Hodgson, V.~Guelpers, R.~Hill and A.~Portelli, \emph{{Split-even approach to the rare kaon decay $K \to \pi \ell^+ \ell^-$}}, \href{https://doi.org/10.22323/1.466.0258}{\emph{PoS} {\bfseries LATTICE2024} (2025) 258} [\href{https://arxiv.org/abs/2501.18358}{{\ttfamily 2501.18358}}].

\bibitem{FLAG:2024oxs}
{\scshape Flavour Lattice Averaging Group (FLAG)} collaboration, \emph{{FLAG Review 2024}},  \href{https://arxiv.org/abs/2411.04268}{{\ttfamily 2411.04268}}.

\bibitem{RQCD:2019jai}
{\scshape RQCD} collaboration, \emph{{Nucleon axial structure from lattice QCD}}, \href{https://doi.org/10.1007/JHEP05(2020)126}{\emph{JHEP} {\bfseries 05} (2020) 126} [\href{https://arxiv.org/abs/1911.13150}{{\ttfamily 1911.13150}}].

\bibitem{He:2021yvm}
J.~He et~al., \emph{{Detailed analysis of excited-state systematics in a lattice QCD calculation of gA}}, \href{https://doi.org/10.1103/PhysRevC.105.065203}{\emph{Phys. Rev. C} {\bfseries 105} (2022) 065203} [\href{https://arxiv.org/abs/2104.05226}{{\ttfamily 2104.05226}}].

\bibitem{Gupta:2021ahb}
R.~Gupta, S.~Park, M.~Hoferichter, E.~Mereghetti, B.~Yoon and T.~Bhattacharya, \emph{{Pion\textendash{}Nucleon Sigma Term from Lattice QCD}}, \href{https://doi.org/10.1103/PhysRevLett.127.242002}{\emph{Phys. Rev. Lett.} {\bfseries 127} (2021) 242002} [\href{https://arxiv.org/abs/2105.12095}{{\ttfamily 2105.12095}}].

\bibitem{Agadjanov:2023efe}
A.~Agadjanov, D.~Djukanovic, G.~von Hippel, H.B.~Meyer, K.~Ottnad and H.~Wittig, \emph{{Nucleon Sigma Terms with Nf=2+1 Flavors of O(a)-Improved Wilson Fermions}}, \href{https://doi.org/10.1103/PhysRevLett.131.261902}{\emph{Phys. Rev. Lett.} {\bfseries 131} (2023) 261902} [\href{https://arxiv.org/abs/2303.08741}{{\ttfamily 2303.08741}}].

\bibitem{Jang:2023zts}
{\scshape Precision Neutron Decay Matrix Elements (PNDME)} collaboration, \emph{{Nucleon isovector axial form factors}}, \href{https://doi.org/10.1103/PhysRevD.109.014503}{\emph{Phys. Rev. D} {\bfseries 109} (2024) 014503} [\href{https://arxiv.org/abs/2305.11330}{{\ttfamily 2305.11330}}].

\bibitem{Alexandrou:2024ozj}
C.~Alexandrou, S.~Bacchio, J.~Finkenrath, C.~Iona, G.~Koutsou, Y.~Li et~al., \emph{{Nucleon charges and $\sigma$-terms in lattice QCD}},  \href{https://arxiv.org/abs/2412.01535}{{\ttfamily 2412.01535}}.

\bibitem{MAIANI1987420}
L.~Maiani, G.~Martinelli, M.~Paciello and B.~Taglienti, \emph{Scalar densities and baryon mass differences in lattice qcd with wilson fermions}, \href{https://doi.org/https://doi.org/10.1016/0550-3213(87)90078-2}{\emph{Nuclear Physics B} {\bfseries 293} (1987) 420}.

\bibitem{GUSKEN1989266}
S.~Güsken, U.~Löw, K.-H.~Mütter, R.~Sommer, A.~Patel and K.~Schilling, \emph{Non-singlet axial vector couplings of the baryon octet in lattice qcd}, \href{https://doi.org/https://doi.org/10.1016/S0370-2693(89)80034-6}{\emph{Physics Letters B} {\bfseries 227} (1989) 266}.

\bibitem{CSSM:2014uyt}
{\scshape CSSM, QCDSF/UKQCD} collaboration, \emph{{Feynman-Hellmann approach to the spin structure of hadrons}}, \href{https://doi.org/10.1103/PhysRevD.90.014510}{\emph{Phys. Rev. D} {\bfseries 90} (2014) 014510} [\href{https://arxiv.org/abs/1405.3019}{{\ttfamily 1405.3019}}].

\bibitem{Bouchard:2016heu}
C.~Bouchard, C.C.~Chang, T.~Kurth, K.~Orginos and A.~Walker-Loud, \emph{{On the Feynman-Hellmann Theorem in Quantum Field Theory and the Calculation of Matrix Elements}}, \href{https://doi.org/10.1103/PhysRevD.96.014504}{\emph{Phys. Rev. D} {\bfseries 96} (2017) 014504} [\href{https://arxiv.org/abs/1612.06963}{{\ttfamily 1612.06963}}].

\bibitem{Savage:2016kon}
M.J.~Savage, P.E.~Shanahan, B.C.~Tiburzi, M.L.~Wagman, F.~Winter, S.R.~Beane et~al., \emph{{Proton-Proton Fusion and Tritium $\beta$ Decay from Lattice Quantum Chromodynamics}}, \href{https://doi.org/10.1103/PhysRevLett.119.062002}{\emph{Phys. Rev. Lett.} {\bfseries 119} (2017) 062002} [\href{https://arxiv.org/abs/1610.04545}{{\ttfamily 1610.04545}}].

\bibitem{QCDSFUKQCDCSSM:2023qlx}
{\scshape QCDSF/UKQCD/CSSM} collaboration, \emph{{Constraining beyond the standard model nucleon isovector charges}}, \href{https://doi.org/10.1103/PhysRevD.108.094511}{\emph{Phys. Rev. D} {\bfseries 108} (2023) 094511} [\href{https://arxiv.org/abs/2304.02866}{{\ttfamily 2304.02866}}].

\bibitem{Barca:2022uhi}
L.~Barca, G.~Bali and S.~Collins, \emph{{Toward N to N\ensuremath{\pi} matrix elements from lattice QCD}}, \href{https://doi.org/10.1103/PhysRevD.107.L051505}{\emph{Phys. Rev. D} {\bfseries 107} (2023) L051505} [\href{https://arxiv.org/abs/2211.12278}{{\ttfamily 2211.12278}}].

\bibitem{Alexandrou:2024tin}
C.~Alexandrou, G.~Koutsou, Y.~Li, M.~Petschlies and F.~Pittler, \emph{{Investigation of pion-nucleon contributions to nucleon matrix elements}}, \href{https://doi.org/10.1103/PhysRevD.110.094514}{\emph{Phys. Rev. D} {\bfseries 110} (2024) 094514} [\href{https://arxiv.org/abs/2408.03893}{{\ttfamily 2408.03893}}].

\bibitem{Barca:2024hrl}
L.~Barca, G.~Bali and S.~Collins, \emph{{Nucleon sigma terms with a variational analysis from Lattice QCD}}, \href{https://doi.org/10.1103/PhysRevD.111.L031505}{\emph{Phys. Rev. D} {\bfseries 111} (2025) L031505} [\href{https://arxiv.org/abs/2412.13138}{{\ttfamily 2412.13138}}].

\bibitem{Tiburzi:2009zp}
B.C.~Tiburzi, \emph{{Time Dependence of Nucleon Correlation Functions in Chiral Perturbation Theory}}, \href{https://doi.org/10.1103/PhysRevD.80.014002}{\emph{Phys. Rev. D} {\bfseries 80} (2009) 014002} [\href{https://arxiv.org/abs/0901.0657}{{\ttfamily 0901.0657}}].

\bibitem{Bar:2016uoj}
O.~B{\"a}r, \emph{{Nucleon-pion-state contribution in lattice calculations of the nucleon charges $g_A,g_T$ and $g_S$}}, \href{https://doi.org/10.1103/PhysRevD.94.054505}{\emph{Phys. Rev. D} {\bfseries 94} (2016) 054505} [\href{https://arxiv.org/abs/1606.09385}{{\ttfamily 1606.09385}}].

\bibitem{Hansen:2016qoz}
M.T.~Hansen and H.B.~Meyer, \emph{{On the effect of excited states in lattice calculations of the nucleon axial charge}}, \href{https://doi.org/10.1016/j.nuclphysb.2017.08.017}{\emph{Nucl. Phys. B} {\bfseries 923} (2017) 558} [\href{https://arxiv.org/abs/1610.03843}{{\ttfamily 1610.03843}}].

\bibitem{Bar:2017kxh}
O.~B{\"a}r, \emph{{Chiral perturbation theory and nucleon\textendash{}pion-state contaminations in lattice QCD}}, \href{https://doi.org/10.1142/S0217751X17300113}{\emph{Int. J. Mod. Phys. A} {\bfseries 32} (2017) 1730011} [\href{https://arxiv.org/abs/1705.02806}{{\ttfamily 1705.02806}}].

\bibitem{Bar:2017gqh}
O.~B{\"a}r, \emph{{Multi-hadron-state contamination in nucleon observables from chiral perturbation theory}}, \href{https://doi.org/10.1051/epjconf/201817501007}{\emph{EPJ Web Conf.} {\bfseries 175} (2018) 01007} [\href{https://arxiv.org/abs/1708.00380}{{\ttfamily 1708.00380}}].

\bibitem{Bar:2018xyi}
O.~B{\"a}r, \emph{{$N\pi$-state contamination in lattice calculations of the nucleon axial form factors}}, \href{https://doi.org/10.1103/PhysRevD.99.054506}{\emph{Phys. Rev. D} {\bfseries 99} (2019) 054506} [\href{https://arxiv.org/abs/1812.09191}{{\ttfamily 1812.09191}}].

\bibitem{Jang:2019vkm}
Y.-C.~Jang, R.~Gupta, B.~Yoon and T.~Bhattacharya, \emph{{Axial Vector Form Factors from Lattice QCD that Satisfy the PCAC Relation}}, \href{https://doi.org/10.1103/PhysRevLett.124.072002}{\emph{Phys. Rev. Lett.} {\bfseries 124} (2020) 072002} [\href{https://arxiv.org/abs/1905.06470}{{\ttfamily 1905.06470}}].

\bibitem{Bar:2019gfx}
O.~B{\"a}r, \emph{{$N\pi$-state contamination in lattice calculations of the nucleon pseudoscalar form factor}}, \href{https://doi.org/10.1103/PhysRevD.100.054507}{\emph{Phys. Rev. D} {\bfseries 100} (2019) 054507} [\href{https://arxiv.org/abs/1906.03652}{{\ttfamily 1906.03652}}].

\bibitem{Bar:2021crj}
O.~B{\"a}r and H.~Colic, \emph{{N\ensuremath{\pi}-state contamination in lattice calculations of the nucleon electromagnetic form factors}}, \href{https://doi.org/10.1103/PhysRevD.103.114514}{\emph{Phys. Rev. D} {\bfseries 103} (2021) 114514} [\href{https://arxiv.org/abs/2104.00329}{{\ttfamily 2104.00329}}].

\bibitem{Luscher:2001up}
M.~L{\"u}scher and P.~Weisz, \emph{{Locality and exponential error reduction in numerical lattice gauge theory}}, \href{https://doi.org/10.1088/1126-6708/2001/09/010}{\emph{JHEP} {\bfseries 09} (2001) 010} [\href{https://arxiv.org/abs/hep-lat/0108014}{{\ttfamily hep-lat/0108014}}].

\bibitem{Gerardin:2023naa}
A.~G\'erardin, W.E.A.~Verplanke, G.~Wang, Z.~Fodor, J.N.~Guenther, L.~Lellouch et~al., \emph{{Lattice calculation of the $\pi^0$, $\eta$ and $\eta^{\prime}$ transition form factors and the hadronic light-by-light contribution to the muon $g-2$}},  \href{https://arxiv.org/abs/2305.04570}{{\ttfamily 2305.04570}}.

\end{thebibliography}\endgroup
